\documentclass[a4paper,10pt,preprint,sort&compress]{elsarticle}
\pdfoutput=1
\usepackage[T1]{fontenc}\usepackage[latin1]{inputenc}
\usepackage{dcolumn,graphicx,color,wrapfig,booktabs,microtype,afterpage} \graphicspath{{Figures/}}
\usepackage[charter,greekuppercase=italicized]{mathdesign}\usepackage{sidecap}
\renewcommand{\figurename}{Fig.}
\renewcommand{\tablename}{Table}
\makeatletter\renewcommand{\fnum@figure}[1]{\textbf{\figurename~\thefigure.}}\makeatother
\makeatletter\renewcommand{\fnum@table}[1]{\textbf{\tablename~\thetable.}}\makeatother

\newcount\hh \newcount\mm
\hh=\time \divide\hh by 60
\mm=\hh \multiply\mm by 60 \mm=-\mm
\advance\mm by \time
\def\now{\number\hh:\ifnum\mm<10{}0\fi\number\mm}

\usepackage[colorlinks,plainpages=false,linkcolor=blue,urlcolor=blue,citecolor=blue,pdfpagemode=UseNone,pdfstartview=FitBH]{hyperref}

\newcommand{\half}{\frac{1}{\protect\raisebox{0.8pt}{\scriptsize 2}}}
\newcommand{\quarter}{\frac{1}{\protect\raisebox{0.8pt}{\scriptsize 4}}}
\newcommand{\threehalf}{\frac{3}{\protect\raisebox{0.8pt}{\scriptsize 2}}}

\journal{Comptes Rendus Physique}

\begin{document}

\makeatletter\renewcommand{\ps@plain}{%
\def\@evenhead{\small\hfill\textit{D.~S.~Inosov~/~Spin fluctuations in ferropnictides~/~Comptes Rendus Physique (preprint)}\hfill\thepage}%
\def\@oddhead{\small\hfill\textit{D.~S.~Inosov~/~Spin fluctuations in ferropnictides~/~Comptes Rendus Physique (preprint)}\hfill\thepage}%
}\makeatother\pagestyle{plain}

\begin{frontmatter}

\title{\flushleft\LARGE\textsf{Spin fluctuations in iron pnictides and chalcogenides:\\ From antiferromagnetism to superconductivity.}}

\author{Dmytro S. Inosov}
\address{Institut f\"ur Festk\"orperphysik, TU Dresden, D-01069 Dresden, Germany}
\ead{dmytro.inosov@tu-dresden.de}

\begin{abstract}
The present article reviews recent experimental investigations of spin dynamics in iron-based superconductors and their parent compounds by means of inelastic neutron scattering. It mainly focuses on the most contemporary developments in this field, pertaining to the observations of magnetic resonant modes in new superconductors, spin anisotropy of low-energy magnetic fluctuations that has now been observed in a wide range of chemical compositions and doping levels, as well as their momentum-space anisotropy incurred by the spin-nematic order. The implications of these new findings for our understanding of the superconducting state, along with the remaining unsettled challenges for neutron spectroscopy, are discussed.
\end{abstract}


\end{frontmatter}


\section{Introduction}

\subsection{A brief history of Fe-based superconductors}

Layered iron pnictides and chalcogenides entered the limelight of modern condensed-matter physics after it was demonstrated in 2008 that doping fluorine into the layered antiferromagnetic (AFM) compound LaFeAsO induces high-temperature superconductivity with a critical temperature, $T_{\rm c}$, of 26\,K \cite{KamiharaWatanabe08}. This discovery shortly preceded the 100$^{\rm th}$ anniversary of the liquefaction of helium, which led to the first observation of superconductivity in mercury by Heike Kamerlingh Onnes in 1911 \cite{KamerlinghOnnes11}, thus initiating a centennial quest for understanding the underlying mechanisms of this remarkable phenomenon.

Within just a few months after the initial discovery, the record $T_{\rm c}$ of iron-based superconductors was raised to 55\,K in a structurally related SmFeAsO$_{1-\delta}$ \cite{RenChe08, RenLu08} or even 56\,K in Gd$_{1-x}$Th$_x$FeAsO \cite{WangLi08}, which remains unsurpassed among bulk non-cuprate superconductors. In single-layer FeSe thin films, even higher $T_{\rm c}$'s up to 100\,K have been observed \cite{GeLiu14}. In the last 5 years, iron-based superconductors with a multitude of various structures and chemical compositions were discovered \cite{Ivanovskii08, IzyumovKurmaev08, RenZhao09, Chu09, WenLi11, HaoYunLei13}, and numerous detailed reviews on the structural diversity and rich physics of this class of materials became available \cite{Sadovskii08, IshidaNakai09, MizuguchiTakano10, AswathyAnooja10, PaglioneGreene10, Johnston10, LumsdenChristianson10review, Stewart11, InosovPark11, HirschfeldKorshunov11, FisherDegiorgi11, DaiHu12, Kordyuk12, KotegawaFujita12, LongWeiQiang13, Dagotto13, DingLin14, Charnukha14, ChenDai14, Mannella14, ChenLin14, Bao15, JohnsonXu15}. It is now common to classify iron-based superconductors into several \emph{structural families}, denoted shortly by the stoichiometric ratios of chemical constituents in their undoped parent compounds, e.g. ``11'' for binary iron-chalcogenide compounds, like FeSe or FeTe, ``111'' for ternary compounds like LiFeAs or NaFeAs, ``1111'' for the originally discovered $R$FeAsO ($R$~=~rare earth), ``122'' for $A$Fe$_2$As$_2$ ($A$~=~alkali metal) with ThCr$_2$Si$_2$-type crystal structure, and others \cite{Stewart11}.

According to the present consensus, metallic character of undoped ferropnictides and weaker electron correlations, as compared to high-$T_{\rm c}$ cuprates \cite{QazilbashHamlin09, TranquadaXu14}, should lead to a qualitatively simpler description of their underlying physics. This is best illustrated by numerous successes of itinerant theoretical models in accurately reproducing or sometimes predicting the energy-momentum structure of the spin-fluctuation spectrum in the normal and superconducting states \cite{KorshunovEremin08, MaierScalapino08, MaierGraser09, KnolleEremin10a, InosovPark10, ParkInosov10, ParkHaule11, RoweKnolle12, FriemelPark12, YinHaule14}. On the other hand, various quantitative complications arise due to the multiband and multiorbital character of the electronic structure \cite{EschrigKoepernik09, AndersenBoeri11, YuSi11, TranquadaXu14}, effects of disorder, magnetic or nonmagnetic impurities \cite{WeberMila12, FernandesVavilov12, LiShen13, InosovFriemel13, GastiasoroHirschfeld13, WangKreisel13}, and the importance of spin-orbit coupling and spin anisotropy \cite{Ye11, YangZheng12, KorshunovTogushova13, CvetkovicVafek13, BorisenkoEvtushinsky14, FernandesVafek14}. Moreover, the much broader variety of crystal structures and chemical compositions among different families of ferropnictides makes it generally unlikely that conclusions drawn for a single material can be straightforwardly generalized for the whole class of compounds. For instance, while the nodeless $s_\pm$ symmetry of the superconducting order parameter has been suggested for some of the most actively studied families with the highest values of $T_{\rm c}$ \cite{MazinSingh08, YaoLi09, ChenTsuei10, HanaguriNiitaka10}, a few other iron-arsenide compounds display nodes in the gap structure, most notably LaFePO \cite{FletcherSerafin09, HicksLippman09, YamashitaNakata09, SutherlandDunn12}, LiFeP \cite{HashimotoKasahara12}, Sr$_2$ScFePO$_3$ \cite{YatesUsman10}, BaFe$_2$(As$_{1-x}$P$_x$)$_2$ \cite{HashimotoYamashita10, ZhangYe12, MorisakiIshii14}, KFe$_2$As$_2$ \cite{HashimotoSerafin10, DongZhou10, KawanoFurukawa11, TaftiJuneauFecteau13} and RbFe$_2$As$_2$ \cite{ZhangWang15}, evidencing a qualitatively different type of order-parameter symmetry, possibly a nodal $s^{\pm}$ \cite{ThomalePlatt11} or $d$-wave \cite{Bang12}. Since KFe$_2$As$_2$ represents the end member of the Ba$_{1-x}$K$_x$Fe$_2$As$_2$ series of superconductors, the presence of nodes in its gap structure implies that the superconducting gap undergoes a qualitative change of symmetry as a function of a continuously tunable parameter, such as doping \cite{Korshunov14} or hydrostatic pressure \cite{TaftiJuneauFecteau13}. Furthermore, in the iron-phosphide systems LaFePO and Sr$_2$ScO$_3$FeP, inelastic neutron scattering (INS) on powder samples revealed no evidence of strong magnetic fluctuations, indicating their reduced importance for superconductivity as opposed to analogous iron-arsenide materials \cite{TaylorEwings13}. In fact, such a notable spread of physical descriptions for different iron-based superconducting materials is actually not surprising in view of the different magnetic ground states of their parent compounds, Fermi-surface topologies, and a broad range of impurity scattering rates.

\subsection{The role of spin fluctuations in the superconducting pairing mechanism}

Long before the discovery of superconductivity in iron pnictides, theories of unconventional Cooper pairing mediated by spin fluctuations \cite{BerkSchrieffer66} were put forward to explain the superconducting mechanisms in high-$T_{\rm c}$ cuprates, heavy electron systems, and two-dimensional (2D) organic compounds \cite{Scalapino99, MoriyaUeda03, Eschrig06, MonthouxPines07}. Ferropnictides presented a new playground for the development of these theoretical models in view of the tendency of superconductivity to be enhanced near the AFM quantum critical point, where the spin fluctuations are most intense \cite{DaiSi09, Mazin10, AbrahamsSi11, WangLee11, YuLi13}. Due to the multiband character of the Fermi surface, supporting intense paramagnetic fluctuations peaked at the zone boundary, a sign-changing order parameter with an $s_\pm$ or $d$-wave symmetry is fostered by these theories \cite{MazinSingh08, YaoLi09, Korshunov14}. It has also been shown early on that the microscopic coexistence of superconductivity with the long-range magnetic order that is found in several families of iron pnictides is only possible if the superconducting
\begin{wrapfigure}[25]{r}{0.465\textwidth}\vspace{-7pt}
\noindent\includegraphics[width=0.465\textwidth]{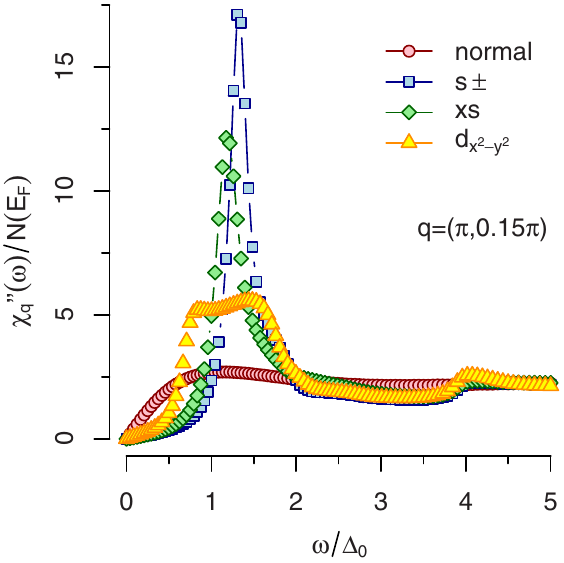}\vspace{-5pt}
\caption{Energy dependence of the dynamic spin susceptibility, $\chi''_\mathbf{q}(\omega)$, calculated within the random phase approximation for the superconducting state, which illustrates the sensitivity of the magnetic resonant mode to the symmetry of the superconducting order parameter. The energy is normalized by the superconducting gap, $\Delta_0$. Reproduced from Ref.\,\citenum{MaierGraser09}, copyright by the American Physical Society.}
\label{Fig:ResonancesTheory}
\end{wrapfigure}
state is $s_\pm$ \cite{FernandesPratt10, VorontsovVavilov10}. An alternative approach that received considerable recent attention is the orbital-fluctuation theory, placing Fe\,$3d$ orbital fluctuations associated with the orthorhombic transition in the driver's seat for superconductivity \cite{KontaniOnari10, OnariKontani10, SaitoOnari10}. These two scenarios are difficult to disentangle, as spin and orbital fluctuations go hand in hand and can not be easily separated either in experiment or in theory \cite{HirschfeldKorshunov11}. The most direct proof of a sign change in the gap structure is expected from future phase-sensitive tests based on combinations of tunnel junctions and point contacts \cite{GolubovMazin13}, analogous to those applied previously to copper oxides \cite{WollmanHarlingen93, TsueiKirtley94, VanHarlingen95, TsueiKirtley00}. Until now, the design of similarly well-controlled phase-sensitive experiments for iron-based superconductors still remains in its infancy. The statistical data obtained so far on the flux quantization in polycrystalline NdFeAsO$_{0.88}$F$_{0.12}$ \cite{ChenTsuei10} provide indirect evidence for the existence of Josephson loops with a $\pi$ phase shift in support of the sign-changing $s_\pm$ gap symmetry. More ``pure'' experiments on individual Josephson junctions with controlled orientation and on a broader range of compounds are in principle possible \cite{GolubovMazin13, LiuTao14}, but still await practical realization.

While there appears to be no single smoking-gun experiment so far that could unequivocally demonstrate the sign change of the order parameter in iron-based superconductors or single out the pairing boson, proponents of spin-fluctuation theories find multiple supporting evidence in spectral signatures of collective magnetic excitations in the electronic structure or, vice versa, of the superconducting gap opening in the spin-fluctuation spectra. The former ones include anomalies in the tunneling spectra \cite{ShanGong12, WangYang12} that match the energy of the neutron spin resonance, observations of ``kinks'' in photoemission data \cite{RichardSato09, EvtushinskyKim11} or quasiparticle-interference patterns \cite{AllanLee15}, and superconductivity-induced infrared optical anomalies \cite{CharnukhaDolgov11} that could originate from strong coupling to a bosonic spectral function resembling that of spin fluctuations. Although these signatures are not direct proof that the corresponding boson is involved in the pairing mechanism, they may contain quantitative information about electron-magnon coupling strength that could potentially facilitate future estimates of $T_{\rm c}$ along the lines developed recently for high-$T_{\rm c}$ cuprates \cite{DahmHinkov09, LeTaconGhiringhelli11}.

The reverse effect of band gap opening on the magnetic excitation spectrum manifests itself in the formation of the magnetic resonant mode, revealed by INS as a redistribution of magnetic spectral weight below $T_{\rm c}$ from the spin-gap region at low energies into a relatively sharp peak centered near or below the gap energy, $2\Delta$. The neutron spin resonance has now been established in most iron-based superconductors \cite{ChristiansonGoremychkin08, LumsdenChristianson09, ChiSchneidewind09, QiuBao09, ZhaoRegnault10, InosovPark10, ShamotoIshikado10, IshikadoNagai11, SatoKawamata11, ParkFriemel11, FriemelLiu12, TaylorEwings12, TaylorSedlmaier13, PriceSu13, ZhangYu13, ZhangLi13, ZhaoRotundu13} with very few exceptions like LiFeAs \cite{TaylorPitcher11, WangWang12, QureshiSteffens12_LiFeAs, QureshiSteffens14}, where the spectral-weight redistribution is much less pronounced and
\begin{wrapfigure}[28]{r}{0.44\textwidth}\vspace{-3pt}
\noindent\includegraphics[width=0.44\textwidth]{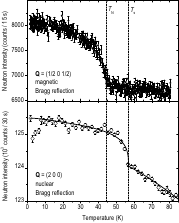}\vspace{-3pt}
\caption{Elastic neutron-scattering measurements of the magnetic (top) and structural (bottom) phase transitions in NaFeAs, demonstrating a relatively broad region between $T_{\rm s}$ and $T_{\rm N}$, attributed to the spin-nematic phase with spontaneously broken C$_4$ symmetry within the $a{\kern-.3pt}b$ plane. Reproduced from Ref.\,\citenum{ParkFriemel12}, copyright by the American Physical Society.}
\label{Fig:NFA-transitions}
\end{wrapfigure}
leaves room for ambiguous interpretations \cite{KnolleZabolotnyy12}. It has been noted early on that the intensity and shape of the magnetic resonance peak in the INS spectrum would sensitively depend on the pairing symmetry \cite{KorshunovEremin08, MaierScalapino08, MaierGraser09}, so that the coherence factor would stabilize a pronounced collective mode only for sign-changing order parameters, as illustrated in Fig.\,\ref{Fig:ResonancesTheory}. Consequently, most authors consider the observation of intense resonant modes in some iron-based superconductors as the leading argument in support of the spin-fluctuation theory \cite{HirschfeldKorshunov11, AkbariEremin09, MaitiKnolle11, KhodasChubukov12, FriemelPark12, ZhangLi13, PandeyChubukov13}, while some others propose alternative explanations for the observed resonance peak that requires no sign change in the gap function \cite{OnariKontani10, OnariKontani11, QureshiSteffens12_LiFeAs}. This controversy places neutron spectroscopy among the key techniques that are expected to shed light on the origin of high-$T_{\rm c}$ superconductivity in ferropnictides, and we will turn to a more detailed discussion of the available experimental data and their possible interpretations \mbox{further on in this review}.

\subsection{Understanding the spin-nematic phase}\label{SubSec:UnderstandingNematic}\enlargethispage{1pt}

As for the Cooper pairing mechanism, either spin \cite{FangYao08, XuMuller08, FernandesBohmer13, NakaiIye13} or orbital \cite{LvWu09, KimJung13, BaekEfremov14} degrees of freedom are also considered responsible for the orthorhombic phase transition in the parent or lightly doped iron arsenides, which usually precedes the onset of static AFM order, leading to the formation of the so-called spin-nematic state in the narrow temperature region between the two transitions \cite{MazinJohannes09, LeeLv13, DavisHirschfeld14, FernandesChubukov14}. Because the AFM and spin-nematic quantum critical points are nearly coincident in the generic phase diagram of iron-based superconductors, figuring out which one of them supplies the ``pairing glue'' still remains a subject of debate, further complicated by our limited understanding of the spin-nematic phase on its own. The origin of nematicity is therefore tightly linked to the dilemma of the Cooper pairing mechanism. A major obstacle \emph{en route} to the experimental clarification of this important problem is the natural twinning of the orthorhombic domains in the crystal, which masks the anisotropic effects in the nematic phase. Because the structural transition in all the compounds studied so far lies below room temperature, the preparation of a single-domain sample is only possible by detwinning the crystal \emph{in situ} \cite{FisherDegiorgi11} using either uniaxial stress \cite{TanatarBlomberg10, ChuAnalytis10, YingWang11, ChuKuo12, KuoShapiro13, KuoFisher14, DuszaLucarelli11, DuszaLucarelli12, NakajimaLiang11, NakajimaIshida12, KimOh11, YiLu11, YiLu12, ZhangHe12, LuPark14} or in-plane magnetic field \cite{ChuAnalytis10prb, XiaoSu10, ZapfStingl14} applied during cooling. The results of these experiments will be reviewed in detail in what follows.

Despite the technical difficulties involved in the mechanical detwinning procedure, several experimental techniques have already succeeded in measuring the single-domain orthorhombic state: mainly x-ray diffraction \cite{TanatarBlomberg10}, electron transport \cite{TanatarBlomberg10, ChuAnalytis10, YingWang11, ChuKuo12, KuoShapiro13, KuoFisher14}, optics \cite{DuszaLucarelli11, DuszaLucarelli12, NakajimaLiang11, NakajimaIshida12}, photoelectron spectroscopy \cite{KimOh11, YiLu11, YiLu12, ZhangHe12} and, most recently, neutron scattering \cite{LuPark14}. However, only few of them could so far distinguish the AFM and spin-nematic transitions clearly in an untwinned crystal. To achieve this, the experiment would have to conform to the following important criteria, which are highly difficult to be realized simultaneously:

\begin{enumerate}
\item \emph{Performing control experiments.} If too much stress is applied to the sample, irreversible changes may occur, such as the formation of cracks or dislocations in the material, leading to anisotropic properties that do not fully disappear even after the external stress is fully removed. In order to exclude such effects, the experimental setup must allow for the release of the stress while keeping other measurement parameters constant. Control experiments in the absence of external stress should be then performed both before and after the experiment, demonstrating that the measured sample properties remain isotropic and consistent in both measurements.
\item \emph{Measuring the twinning ratio.} In order to demonstrate that the sample remains fully untwinned, it is desirable to measure the twinning ratio directly during the experiment (e.g. by diffraction or polarized optical imaging) or to perform preliminary tests with the same sample environment to ensure that the twinning ratio is sufficiently low and does not change in the whole accessible temperature range, as it could hinder accurate measurements.
\item \emph{Using compounds with distinct structural and magnetic phase transitions.} To be able to access the narrow region of the spin-nematic phase, the measured compound should be characterized by a large difference between the structural and magnetic phase transition temperatures, $T_{\rm s}-T_{\rm N}$. This rules out, for instance, the stoichiometric parent compounds of the ``122'' family, where the two phase transitions nearly coincide. So far, the largest differences between the two transition temperatures were reported in SrFeAsF with $T_{\rm s}-T_{\rm N}\approx50$\,K \cite{BakerFranke09, LiFang11}, Ca$_{10}$(Pt$_3$As$_8$)(FeAs)$_{10}$ with $T_{\rm s}-T_{\rm N}\approx14$\,K \cite{SapkotaTucker14}, and Na$_{1-\delta}$FeAs with $T_{\rm s}-T_{\rm N}\approx12$\,K \cite{ChenHu09, LiCruz09, ParkerSmith10, WrightLancaster12, ParkFriemel12} (see Fig.~\ref{Fig:NFA-transitions}). These materials are air sensitive and therefore more difficult to handle as compared to the ``122'' compounds. The detwinning experiments on such systems are so far limited to angle-resolved photoemission studies of NaFeAs under uniaxial pressure \cite{YiLu12, ZhangHe12}, in which distinct signatures of orbital-dependent electronic reconstruction were observed across both $T_{\rm s}$ and $T_{\rm N}$. There is also a lot of current interest in understanding the spin-nematic phase in FeSe$_{1-\delta}$ \cite{McQueenWilliams09, BoehmerHardy13, ShimojimaSuzuki14, NakayamaMiyata14, WatsonKim15} that extends from $T_{\rm s}=90$\,K down to zero temperature with no associated static magnetic order, unless a hydrostatic pressure above 0.8\,GPa is applied to the sample \cite{BendeleAmato10, BendeleIchsanow12}.
\item \emph{Releasing external stress at low temperatures.} It is now well known that uniaxial stress induces strong anisotropy effects in layered iron pnictides even at temperatures above $T_{\rm s}$, as evidenced, for example, by anomalously large and anisotropic elastoresistance in comparison to simple metals \cite{ChuKuo12, KuoShapiro13}. This results in smearing of the structural phase transition, so that $T_{\rm s}$ and $T_{\rm N}$ finally merge and the nematic state is no longer well defined \cite{FisherDegiorgi11}. The uniaxial pressure can also influence the characteristic transition temperatures \cite{DhitalYamani12, DhitalHogan14, QinDong15} or even introduce a decoupling between the onsets of the orthorhombic distortion and antiferromagnetism \cite{DhitalHogan14}. At the same time, high stress is required to achieve low twinning ratios, especially in large crystals. Moreover, since the spin-nematic state is understood as a phase with spontaneously broken 4-fold rotational symmetry, a ``clean'' experiment has to be performed under isotropic external conditions, in particular under zero stress, as was already emphasized earlier \cite{Charnukha14}. This inevitably leads to the conclusion that an experiment carried out in a detwinning device supplying constant uniaxial pressure can only provide useful information about the anisotropy below $T_{\rm N}$, but is not suitable for investigations of the spin-nematic phase between $T_{\rm N}$ and $T_{\rm s}$. Instead, devices that allow for the release of the uniaxial pressure in a controlled way \emph{in situ} during the experiment have to be employed. Accurate calibration is then required to ensure that isotropic conditions are indeed restored after the detwinning procedure. The existing devices with such possibilities are usually piezoelectrically driven, so that remote control of the exerted stress becomes possible \cite{HicksBarber14}. The reported measurements are so far restricted to the anisotropy of resistivity \cite{ChuKuo12, KuoShapiro13, KuoFisher14} and optical response \cite{MirriDusza14}, and were performed exclusively on the parent and lightly doped `122' compounds, where the spin-nematic state is difficult to resolve.
\end{enumerate}

An alternative method of probing the single-domain nematic state is to reduce the sample to the dimensions comparable with the size of a single naturally formed orthorhombic twin domain \cite{KasaharaShi12} or to use local probes with atomic resolution, such as scanning tunneling spectroscopy \cite{AllanChuang13, RosenthalAndrade14, CaiRuan14}. One should not forget, however, that truly isotropic conditions can not be achieved in a local-probe experiment on a twinned crystal because of the internal strains accumulated in the sample during its preparation, which can be partially released by annealing, and the unavoidable stresses exerted by neighboring domains below the structural transition. As a result, the material remains locally anisotropic far above $T_{\rm s}$, resembling experimental results under constantly applied stress \cite{RosenthalAndrade14}.

\section{Latest progress in material synthesis}

\begin{wrapfigure}[19]{r}{0.35\textwidth}\vspace{-35pt}
\noindent\includegraphics[width=0.35\textwidth]{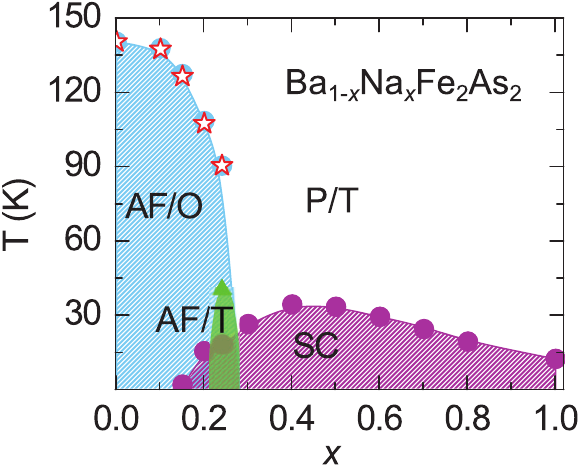}
\caption{The phase diagram of Ba$_{1-x}$Na$_x$Fe$_2$As$_2$, which exhibits an additional tetragonal phase (AF/T) inside the usual orthorhombic AFM dome (AF/O). Reproduced from Ref.\,\citenum{AvciAllred13}, copyright by the American Physical Society.}
\label{Fig:BNFA}
\end{wrapfigure}

The quest for new families of ferropnictides and the optimization of their synthesis in terms of both quality and the size of single crystals available for experimental investigations remains at the forefront of current research. In recent years, these efforts persisted along several major directions, enabling a significant progress in our understanding of the ever increasing number of structurally different compounds across a wide range of doping levels and other control parameters. Here we take a look at some of the important milestones along this route, mostly relevant for neutron spectroscopy.

\begin{figure*}[b!]\vspace{-4pt}
\begin{center}
\includegraphics[width=0.9\textwidth]{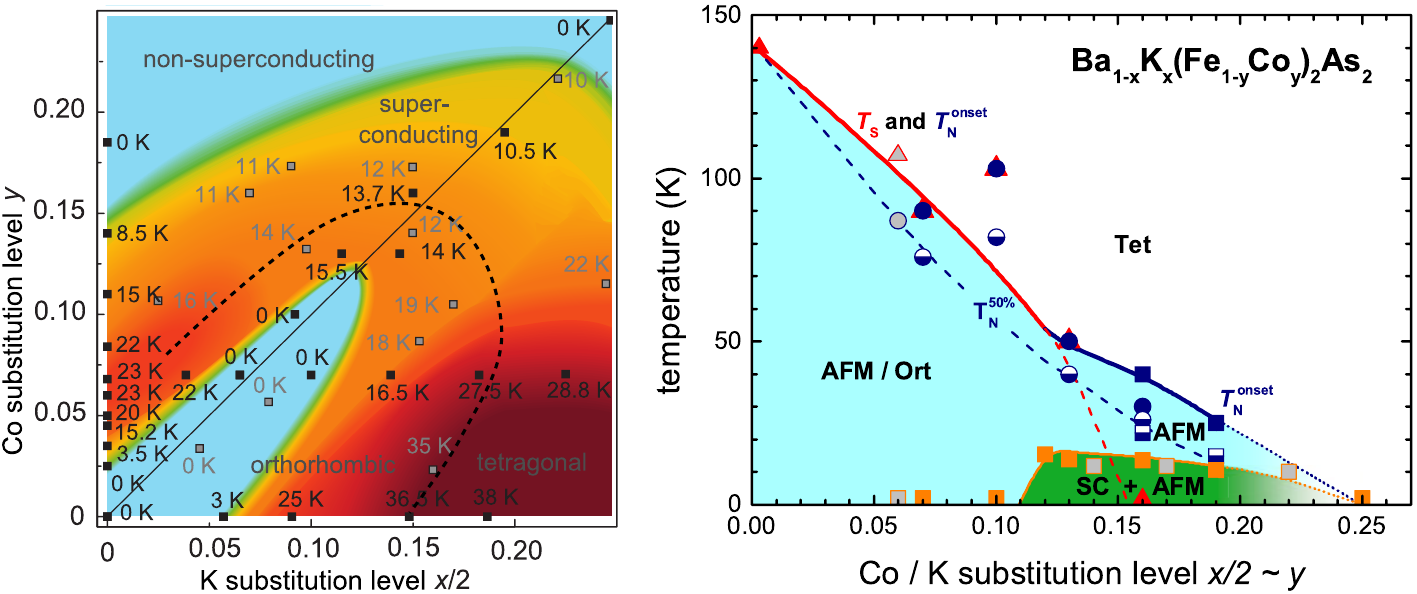}\vspace{-6pt}
\end{center}
\caption{Left: The ground state of Ba$_{1-x}$K$_x$(Fe$_{1-y}$Co$_y$)$_2$As$_2$ vs. dopant concentrations. Blue color represents the nonsuperconducting ground state, the colors from yellow ($T_{\rm c}\approx10$\,K) to dark red ($T_{\rm c}\approx40$\,K) correspond to the superconducting state. Right: The composition-temperature phase diagram of charge-compensated Ba$_{1-x}$K$_x$(Fe$_{1-y}$Co$_y$)$_2$As$_2$ along the $x/2 \approx y$ diagonal, showing the coexistence of the AFM state with superconductivity for $0.10 \lesssim y \lesssim 0.25$. Reproduced from Ref.~\citenum{GoltzZinth14}, copyright by the American Physical Society.\vspace{-6pt}}
\label{Fig:Codoping}
\end{figure*}

\subsection{Moving away from the optimal doping}

While early investigations of iron-pnictide superconductors focused mainly on optimally doped compounds with the highest values of $T_{\rm c}$ and their stoichiometric parents, in recent years the controlled growth of doped single crystals in a broad range of chemical compositions made measurements of highly over- and underdoped samples possible, in particular using neutron-scattering methods. For instance, a detailed study of the spin-fluctuation spectrum in electron-doped Ba-122 systems was performed using several coaligned arrays of under- and overdoped BaFe$_{2-x}$Ni$_x$As$_2$ single crystals with the highest mass up to 45~grams, spanning a broad doping range across the superconducting dome and beyond \cite{LuoYamani12, LuoLu13, WangZhang13}. On the hole-doped side of the phase diagram, where superconductivity persists up to 100\% doping level in Ba$_{1-x}$K$_x$Fe$_2$As$_2$ \cite{AvciChmaissem12}, large single-crystalline samples were so far available for neutron-scattering experiments only near the optimal doping ($x\approx0.33$, up to 10~grams) \cite{ZhangWang11, ZhangLiu13} and for KFe$_2$As$_2$ ($x=1$, up to 3~grams) \cite{LeeKihou11, KawanoFurukawa11, WangZhang13}, while in the intermediate doping range, INS measurements were reported only on powder samples with various potassium concentrations \cite{CastellanRosenkranz11}. Sizable single crystals have also been obtained for the alternatively hole-doped Ba$_{1-x}$Na$_x$Fe$_2$As$_2$ \cite{AswarthamAbdelHafiez12}, in which the sodium ions are significantly smaller than barium ions. While superconductivity in Na- and K-doped compounds behaves similarly \cite{AvciAllred13}, an additional tetragonal magnetic phase has been found upon Na doping inside the usual AFM dome \cite{AvciChmaissem14} (see Fig.\,\ref{Fig:BNFA}) that could be recently connected using neutron diffraction to a reorientation of spins along the $\mathbf{c}$ axis \cite{WasserSchneidewind15}. This material therefore represents a promising new candidate for future spectroscopic investigations, in which new influences of spin orientation and spin-fluctuation anisotropy on superconductivity could be revealed.

Iron pnictides with structures other than ``122'' still remain much less studied by neutron spectroscopy, mostly due to the difficulties in obtaining large enough single crystals, their sensitivity to air \cite{FriemelOhl13}, or both. Nevertheless, considerable attention was paid recently to the parent compounds of NaFeAs \cite{ParkFriemel12, SongRegnault13, ZhangSong14} and LiFeAs \cite{WangWang12, QureshiSteffens12_LiFeAs, QureshiSteffens14} with the ``111''-type structure, as well as some of their doped derivatives, most notably NaFe$_{1-x}$Co$_x$As with a sample mass of 5.5\,g \cite{ZhangLi13}. Another group succeeded in the self-flux growth of sizeable Co- and Rh-doped Na$_{1-\delta}$FeAs single crystals with different dopant concentrations \cite{SteckelRoslova15}. The first INS measurements of spin-wave excitations in the parent La-1111 oxypnictide, performed on an assembly of many small single crystals with a total mass of 600\,mg, also appeared recently \cite{RamazanogluLamsal13}. In superconducting ``1111'' compounds, neutron-spectroscopy experiments are still restricted only to polycrystalline samples of the superconducting LaFeAsO$_{1-\delta}$F$_\delta$ ($T_{\rm c}\leq29$\,K) \cite{ShamotoIshikado10, YamaniRyan12, ShamotoIshikado12} and optimally doped CaFe$_{1-x}$Co$_{x}$AsF ($T_{\rm c}=22$\,K) \cite{PriceSu13}. On the other hand, systematic doping-dependence studies of spin dynamics in different iron-pnictide families were undertaken by other methods, in particular by nuclear-magnetic-resonance (NMR) spectroscopy, as reviewed recently in Ref.\,\citenum{LongWeiQiang13}.

\subsection{Codoping}

Whenever chemical substitution is used to tune physical properties of a system, such as superconductivity, distinguishing the effects of electronic doping from those of impurity scattering or chemical pressure, inevitably associated with the random introduction of dopant atoms, always represents a challenge. To address this problem, some iron-pnictide compounds have been simultaneously codoped with varying concentrations of two elements to maintain a constant charge carrier density while varying the impurity concentration. The most systematic studies of this kind were performed on the Ba-122 system codoped with Co and K \cite{SuzukiOhgushi10, GoltzZinth14}, demonstrating that superconductivity with a $T_{\rm c}$ up to 15.5\,K can be induced even in charge-compensated Ba$_{1-x}$K$_x$(Fe$_{1-y}$Co$_y$)$_2$As$_2$ ($x/2 \approx y$) in coexistence with static AFM order. Figure \ref{Fig:Codoping} summarizes these results in a phase diagram, illustrating the dependence of $T_{\rm c}$ on dopant concentrations and the suppression of the AFM order along the $x/2 = y$ line of charge compensation.

\begin{wrapfigure}[19]{r}{0.5\textwidth}\vspace{-19pt}
\noindent\includegraphics[width=0.5\textwidth]{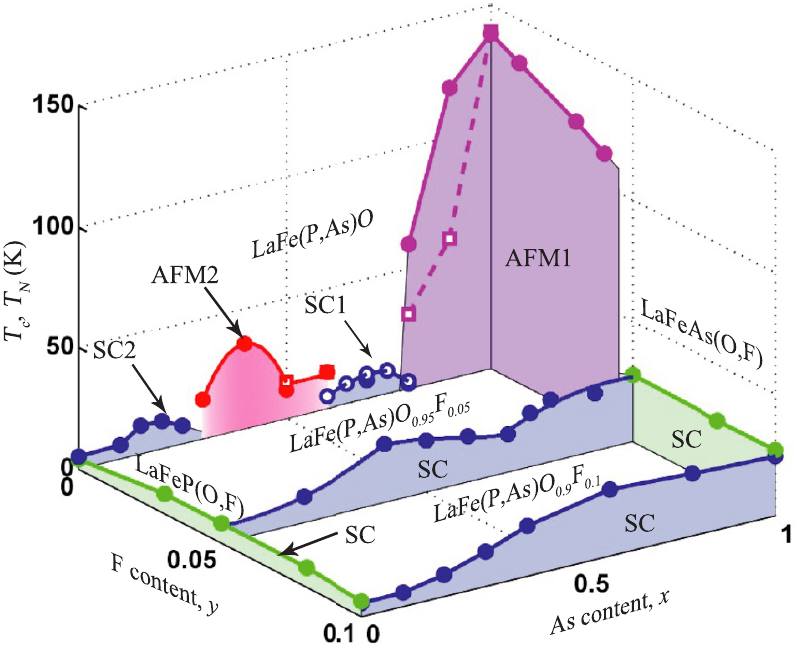}
\caption{The phase diagram of LaFe(As$_{1-x}$P$_x$)O$_{1-y}$F$_y$. Reproduced from Ref.\,\citenum{LaiTakemori14}, copyright by the American Physical Society.}
\label{Fig:Codoping-1111}
\end{wrapfigure}

Another remarkable example is given by the La-1111 compound codoped with fluorine on the oxygen site and phosphorus on the arsenic site \cite{MiyasakaTakemori13, LaiTakemori14}. At zero fluorine concentration, the phase diagram of LaFe(As$_{1-x}$P$_x$)O exhibits a novel AFM phase (with a so far unclarified magnetic structure) for $0.4\leq x\leq 0.7$ with maximal N{\'e}el temperature of 35\,K that intervenes between two superconducting domes \cite{MukudaEngetsu14}. Upon weak fluorine doping, this intermediate phase is suppressed, resulting in a two-humped superconducting dome as seen in the 2D phase diagram shown in Fig.\,\ref{Fig:Codoping-1111}. This suggests the presence of an unusual AFM quantum critical point at some intermediate doping level $0 < y_0 < 0.05$ and a certain phosphorus concentration, $x_{\rm 0}$, such that either increase or decrease in $x$ with respect to $x_0$ should result in an enhancement of $T_{\rm c}$.

Introduction of small amounts of Mn impurities into the superconducting hole-doped Ba$_{1-x}$K$_x$Fe$_2$As$_2$ \cite{ChengShen10, LiGuo12}, electron-doped LaFeAsO$_{1-x}$F$_x$ \cite{HammerathBonfa14} or Ba(Fe$_{1-x}$Co$_x$)$_2$As$_2$ \cite{LeBoeufTexier14}, and isovalently doped BaFe$_2$(As$_{1-x}$P$_x$)$_2$ \cite{LeBoeufTexier14} was also used to demonstrate their strong poisoning effect on superconductivity, associated with the sizable magnetic moments localized on the Mn sites \cite{TexierLaplace12, InosovFriemel13, LeBoeufTexier14}. Thus, in LaFe$_{1-y}$Mn$_y$AsO$_{0.89}$F$_{0.11}$ as little as 0.2\% of Mn is sufficient to completely suppress superconductivity \cite{HammerathBonfa14}, while in Ba$_{0.5}$K$_{0.5}$(Fe$_{1-y}$Mn$_y$)$_2$As$_2$ this critical concentration is much higher and amounts to $\sim$\,5\% \cite{ChengShen10, LiGuo12}, which is nevertheless several times below the typical values for other transition-metal dopants (e.g. Cu, Ni, Zn, Co or Ru) \cite{LiGuo12}. Intermediate values of critical Mn concentrations ($\sim$\,2--3\%) are found in Co- and P-doped BaFe$_2$As$_2$ \cite{LeBoeufTexier14}. Simultaneous substitution of Co and Mn on the iron site was also employed recently to directly demonstrate the localization of induced Mn holes, leading to the absence of any measurable hole doping effect by Mn in angle-resolved photoemission \cite{RienksWolf13}.

\subsection{Iron-chalcogenide superconductors}\label{SubSec:IronChalcogenides}

Single crystals of iron chalcogenides with the ``11''-type structure, such as Fe$_{1+\delta}$Te$_{1-x}$Se$_x$ \cite{MizuguchiTakano10}, along with alkali-metal-intercalated $A_x$Fe$_{2-y}$Se$_2$ ($A$\,=\,K, Rb, Cs, Tl), represent perhaps the second most popular class of iron-based superconductors for neutron scattering investigations after ``122'' compounds. At ambient pressure, the phase diagram of FeTe$_{1-x}$Se$_x$ \cite{DongWang11} exhibits a broad superconducting dome extending from $x\approx0.05$ up to the end member compound FeSe ($T_{\rm c}\approx8.5$\,K) \cite{KawasakiDeguchi12, Singh12, McQueenHuang09}, with a broad maximum around $x=0.5$ reaching $T_{\rm c}^{\rm max}\approx15$\,K. In addition, it was noted that superconductivity in iron selenides is highly sensitive to off-stoichiometry of the Fe content, $\delta$, adding another dimension to the phase diagram \cite{McQueenHuang09, LumsdenChristianson10, StockRodriguez11, TsyrulinViennois12, DucatmanFernandes14}. Initially, a neutron spin resonance was observed in nearly optimally doped FeTe$_{0.6}$Se$_{0.4}$ \cite{QiuBao09}, yet further INS investigations \cite{LumsdenChristianson10, BabkevichBendele10, BabkevichRoessli11, ZaliznyakXu11, StockRodriguez11, ChiRodriguezRivera11, TsyrulinViennois12, ChristiansonLumsden13, StockRodriguez14} also focused on samples with both higher and lower Se concentrations in the range $0 \leq x \leq 0.5$, as well as with different interstitial Fe content, $0.01 \leq \delta \leq 0.14$. Even more recently, substitution of small amounts of Ni or Cu on the iron site was used as an alternative way of tuning superconductivity in these systems \cite{WenLi13, XuWen14}.

The possibility to enhance $T_{\rm c}$ in layered iron selenides by intercalating alkali-metal elements \cite{KrztonMaziopa11, GuoJin10, WangYing11, TsurkanDeisenhofer11, LiuLi12, PengLiu13} or molecular spacer layers \cite{KrztonMaziopa12, ScheidtHathwar12, BurrardLucasFree13, YingChen13, SedlmaierCassidy14} has fostered a new wave of research activity, in which neutron spectroscopy also played a central role. Most commonly available $A_x$Fe$_{2-y}$Se$_2$ compounds \cite{Dagotto13} can be synthesized in the form of large single crystals with very similar values of $T_{\rm c}$ between 27\,K for $A$\,=\,Cs \cite{KrztonMaziopa11} and $\sim$\,32\,K for $A$\,=\,K,\,Rb \cite{GuoJin10, WangYing11, TsurkanDeisenhofer11, LiuLi12, PengLiu13}. However, these materials are infamous for the ubiquitous phase separation into a matrix consisting of non-superconducting vacancy-ordered insulating AFM phase and an intergrown metallic minority phase with at most 10--14\% volume fraction \cite{KsenofontovWortmann11, ShermadiniLuetkens12, CharnukhaCvitkovic12, PomjakushinKrztonMaziopa12}, forming a complex interconnected network of metallic precipitates \cite{CharnukhaDeisenhofer12, LiuXing12, XieYin13, DingFang13, SpellerDudin14, SunWang14} that becomes superconducting below $T_{\rm c}$. It took several years and a combination of modern x-ray and neutron diffraction \cite{RicciPoccia11, ZavalijBao11, PomjakushinKrztonMaziopa12, BosakSvitlyk12, ShoemakerChung12, ZhaoCao12, CarrLouca14}, atomically resolved microscopy \cite{WangSong11, SongWang11, KazakovAbakumov11, LiDing12, CaiYe12, YuanDong12}, optical spectroscopy \cite{CharnukhaDeisenhofer12, HomesXu12, CharnukhaCvitkovic12}, Raman scattering \cite{ZhangLiu12}, photoemission \cite{WangQian11, QianWang11, ZhaoMou11, ZhangYang11, MouLiu11, ChenXu11}, muon-spin relaxation ($\mu$SR) \cite{ShermadiniLuetkens12, CharnukhaCvitkovic12}, NMR \cite{TexierDeisenhofer12, KotegawaTomita12}, M\"ossbauer \cite{KsenofontovWortmann11, StadnikWang13}, and neutron spectroscopy \cite{FriemelPark12, FriemelLiu12, TaylorEwings12} investigations before the true chemical, electronic and microstructural composition of the superconducting phase could be established. According to the present consensus, this phase has an alkali-deficient $A_x$Fe$_2$Se$_2$ composition with vacancy-free FeSe layers, while estimates of the actual amount of dopant atoms still vary widely in the range $0.3 \leq x \leq 0.6$, depending on the employed experimental method \cite{PomjakushinKrztonMaziopa12, FriemelLiu12, TexierDeisenhofer12, CarrLouca14}. The most recently discovered iron selenides containing more complex molecular spacer layers, such as Li$_x$(C$_5$H$_5$N)$_y$Fe$_2$Se$_2$ \cite{KrztonMaziopa12}, (Li/K)$_x$(NH$_2$)$_y$(NH$_3$)$_z$Fe$_2$Se$_2$ \cite{ScheidtHathwar12, BurrardLucasFree13, YingChen13, SedlmaierCassidy14},
\begin{wrapfigure}[17]{l}{0.455\textwidth}\vspace{-7pt}
\noindent\includegraphics[width=0.455\textwidth]{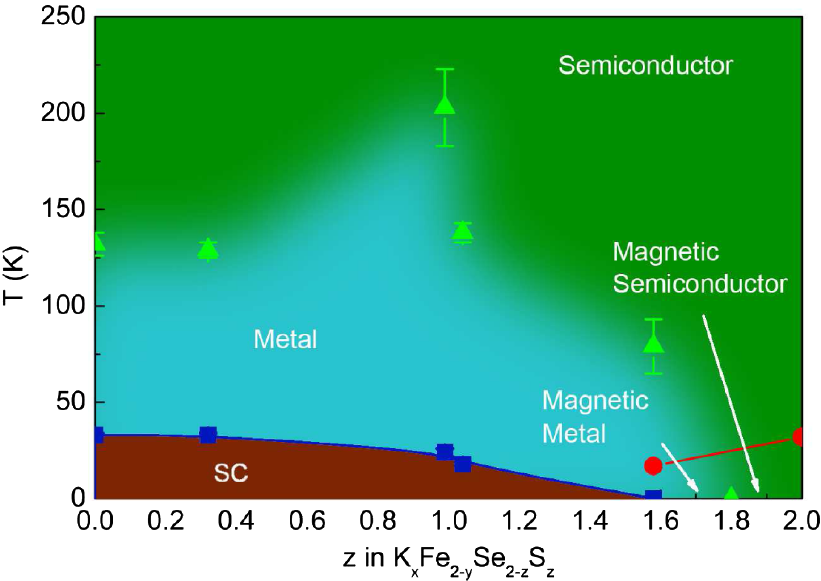}\vspace{-3pt}
\caption{The phase diagram of K$_x$Fe$_{2-y}$Se$_{2-z}$S$_z$, showing a gradual suppression of $T_{\rm c}$ in K$_x$Fe$_{2-y}$Se$_2$ by isovalent sulfur substitution towards the magnetic semiconductor K$_x$Fe$_{2-y}$S$_2$. Reproduced from Ref.\,\citenum{LeiAbeykoon11prl}, copyright by the American Physical Society.}
\label{Fig:FeSeS}
\end{wrapfigure}
or electrically neutral LiFeO$_2$ layers \cite{LuWang14}, along with those intercalated by smaller-sized alkali metals (Li, Na) \cite{YingChen12, WangYing14}, offered a significant improvement of $T_{\rm c}$ up to the record of 46\,K. However, these new materials are so far only available in polycrystalline form, and their superconducting volume fractions are typically low, as can be judged by the low shielding fractions in magnetization measurements. Nevertheless, a recent powder INS measurement on specially prepared deuterated sample of Li$_x$(ND$_2$)$_y$(ND$_3$)$_{1-y}$Fe$_2$Se$_2$ \cite{TaylorSedlmaier13} offered convincing evidence for a resonant enhancement of spectral weight below $T_{\rm c}$ at an unusual incommensurate position that differs from the $(\piup,\piup/2)$ wave vector found in all alkali-metal iron selenides studied this far \cite{ParkFriemel11, FriemelPark12, FriemelLiu12, TaylorEwings12}.

Numerous attempts have been undertaken to continuously control $T_{\rm c}$ in alkali-metal iron selenides by chemical doping. Varying the nominal alkali-metal content typically results in a nearly rectangular superconducting dome with no significant change in $T_{\rm c}$ as a function of nominal composition or Fe valence \cite{TsurkanDeisenhofer11, YanZhang11}. Substitution of various transition metals on the Fe site was only partially successful, as some of them (Cu, Mn, Zn) failed to penetrate the vacancy-free metallic phase and instead showed a tendency to the formation of foreign non-superconducting phases embedded in the sample \cite{SunWang14}, despite a weak enhancement of superconducting properties upon 1--6\% Mn substitution as compared to the non-doped crystals \cite{LiChen13, SunWang14, LiChen14}. Only Ni, Co and Cr atoms could be homogenously doped into the superconducting stripes of the phase-separated K$_x$Fe$_{1-y}$Se$_2$ system, leading to a rapid suppression of $T_{\rm c}$ by just a few percent of the dopant atoms \cite{SunWang14}. A much more gradual and accurate control of $T_{\rm c}$ could be achieved by replacing Se with S \cite{LeiAbeykoon11prl, LeiAbeykoon11prb, TorchettiImai12, ToulemondeSantosCottin13, CarlssonSantosCottin14} or Te \cite{OzakiTakeya13, LiuLi14, WangRyu14}, resulting in phase diagrams like the one presented in Fig.\,\ref{Fig:FeSeS}, continuously connecting the superconducting K$_x$Fe$_{2-y}$Se$_2$ with the semiconducting end member compound, K$_x$Fe$_{2-y}$S$_2$. So far, there are no reports of INS measurements performed on these materials, and therefore the influence of Se-site substitution on the spectrum of spin fluctuations in iron selenides still remains open for future investigations.

\subsection{Iron-platinum-arsenide superconductors}\enlargethispage{2pt}

\begin{wrapfigure}[36]{r}{0.41\textwidth}\vspace{-24pt}
\noindent\includegraphics[width=0.42\textwidth]{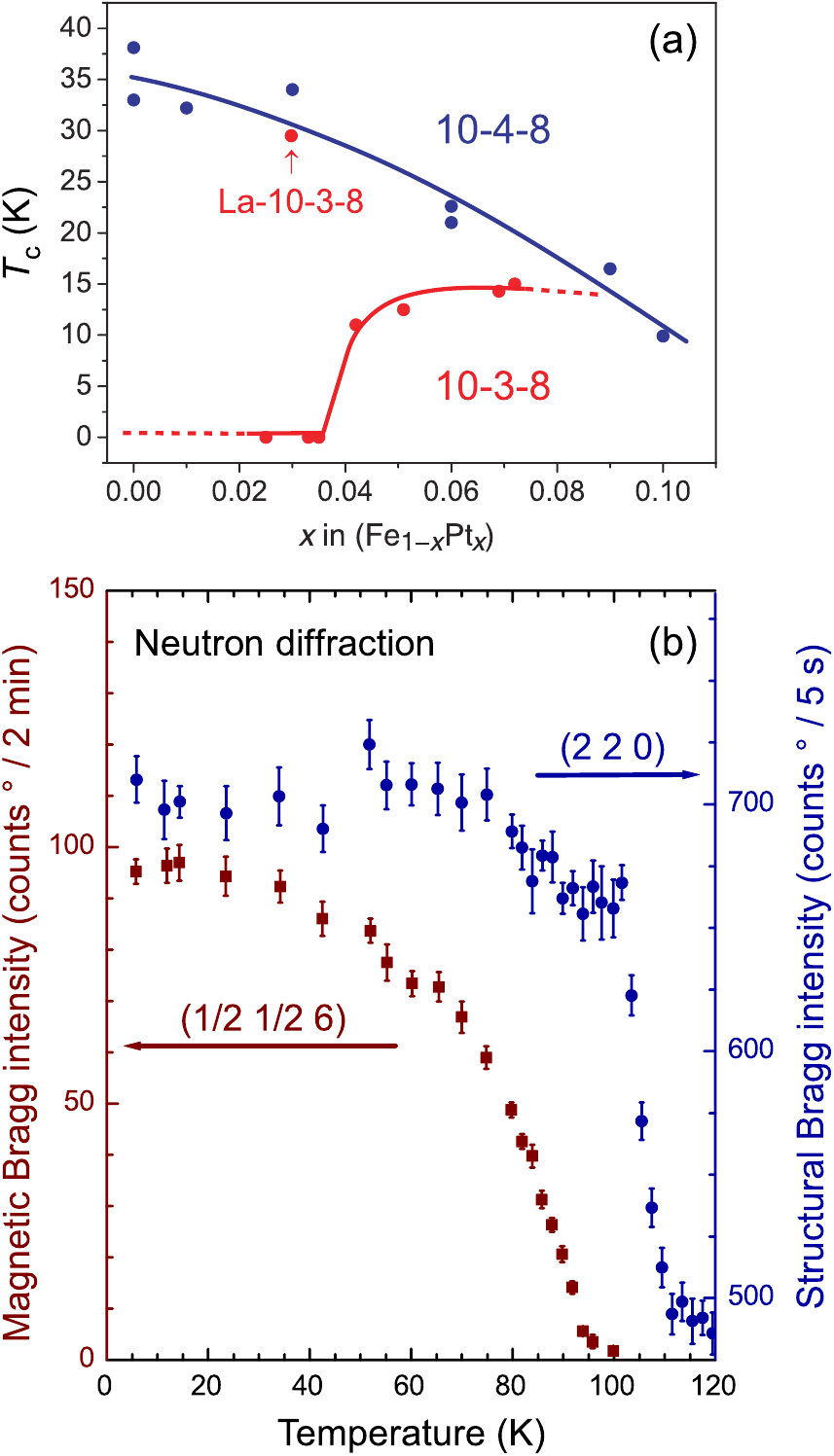}\vspace{-3.5pt}
\caption{(a)~Superconducting transition temperature in ``10-3-8'' and ``10-4-8'' iron arsenides as a function of Pt doping on the Fe site \cite{SturzerDerondeau12}. The point indicated by the arrow shows $T_{\rm c}$ of a La-doped 10-3-8 compound. (b)~Structural and magnetic phase transitions, measured by neutron diffraction on the same ``10-3-8'' single crystal close to the stoichiometric composition \cite{SapkotaTucker14}. Copyright by the American Physical Society.}
\label{Fig:1038-transitions}
\end{wrapfigure}

A few years ago, two new families of iron-arsenide superconductors containing platinum-arsenide intermediary layers, (CaFe$_{1-x}$Pt$_x$As)$_{10}$Pt$_3$As$_8$ and (CaFe$_{1-x}$Pt$_x$As)$_{10}$Pt$_4$As$_8$ (termed ``10-3-8'' and ``10-4-8'', respectively), have been discovered \cite{LohnertSturzer11, NiAllred11, KakiyaKudo11, NoharaKakiya12}. An interesting peculiarity of these systems is that the Pt$_{n}$As$_{8}$ spacer layer ($n=3$ or 4) is metallic, which leads to a peculiar electronic structure with two types of quasi-2D Fermi surfaces originating from Fe\,$d$ and Pt\,$d$ bands \cite{NeupaneLiu12, NakamuraMachida13, Berlijn14}. Apart from that, the ``10-3-8'' family represents a rare example of unconventional superconductors with a low-symmetry triclinic crystal structure. The structural and physical properties of these new materials still represent a subject of intense current investigations \cite{ChoTanatar12, KimRonning12, SturzerFriederichs13, TamegaiDing13, WatsonMcCollam14, ZhouKoutroulakis13, ThirupathaiahSturzer13, XiangLuo12, NiStraszheim13}. Depending on the amount of Pt in the spacer layer, qualitatively different ground-state properties and distinctive behavior of $T_{\rm c}$ have been reported \cite{SturzerDerondeau12}: While the undoped ``10-3-8'' compound is not superconducting and only displays superconductivity upon chemical doping \cite{XiangLuo12, NiStraszheim13, KimSturzer13, SturzerDerondeau14}, the optimal $T_{\rm c} \approx 35$\,K in the related ``10-4-8'' family is reached already in the stoichiometric parent phase and can only be suppressed by doping \cite{NiAllred11, SturzerDerondeau12} [see Fig.\,\ref{Fig:1038-transitions}\,(a)]. Nevertheless, the nearly identical critical temperatures for the parent ``10-4-8'' and optimally rare-earth doped ``10-3-8'' compounds \cite{NiStraszheim13, SturzerDerondeau14} reveal their close relationship despite the structural differences.

Platinum atoms, which are already present in the intermediary layers, can also act as dopants if substituted on the Fe site. This complicates the synthesis of single crystals with the stoichiometric composition, as accidental Pt doping is difficult to avoid. As a result, even the best single crystals of the ``10-3-8'' parent compound obtained so far are characterized by a somewhat lower N\'{e}el temperature ($T_{\rm N}\approx95$\,K) \cite{SapkotaTucker14} as compared to polycrystalline samples ($T_{\rm N}\approx100-120$\,K) \cite{ZhouKoutroulakis13, SturzerFriederichs13}. X-ray diffraction measurements have indicated that the structural transition temperature, $T_{\rm s}\approx120$\,K, considerably exceeds $T_{\rm N}\approx95$\,K in the undoped ``10-3-8'' system \cite{SturzerFriederichs13}. This was more recently confirmed by elastic neutron scattering measurements, where both phase transitions could be measured on the same single-crystalline sample \cite{SapkotaTucker14}, evidencing a relatively high difference, $T_{\rm s}-T_{\rm N}\approx14$\,K, between the structural transition temperature and the onset of stripelike AFM order. As already mentioned earlier, this observation makes ``10-3-8'' compounds particularly promising for future experimental investigations of the spin-nematic phase.

Various ways of inducing superconductivity in the ``\mbox{10-3-8}'' compounds have been explored, including the substitution of platinum \cite{NiAllred11, LohnertSturzer11, KakiyaKudo11, XiangLuo12, ChoTanatar12, KimRonning12, SturzerDerondeau12, SturzerFriederichs13, NoharaKakiya12, TamegaiDing13, WatsonMcCollam14} or other transition metals \cite{SturzerKessler14} on the Fe site, trivalent-metal doping (Y, La--Sm, Gd--Lu) on the Ca site \cite{NiStraszheim13, KimSturzer13, SturzerDerondeau14} or the application of hydrostatic pressure \cite{GaoSun14}. The resulting phase diagrams resemble those of most other iron-arsenide superconductors, where chemical doping suppresses the spin-density-wave (SDW) ground state of the stoichiometric compound and stabilizes a superconducting dome around the AFM quantum critical point \cite{Johnston10}. The superconducting phase in platinum-iron-arsenides is characterized by a large anisotropy of the upper critical fields \cite{NiAllred11, XiangLuo12, NiStraszheim13, NoharaKakiya12, WatsonMcCollam14}, reaching $\gamma_{H} = H^{\parallel}_{\rm c2}(T)/ H^{\bot}_{\rm c2}(T)\approx10$ upon approaching $T_{\rm c}$ \cite{NiAllred11, XiangLuo12, WatsonMcCollam14}, and low-temperature coherence lengths of $\xi_{\parallel}(0)=50$\,\AA\ and $\xi_{\bot}(0)=12$\,\AA\ \cite{NiAllred11} that are comparable to typical values in other iron-based superconductors \cite{Johnston10}. The ratio of the heat capacity jump to the normal-state Sommerfeld coefficient, $\Delta C/\gamma_{\rm n}T_{\rm c}$, was reported for the La-doped ``10-3-8'' compound to be 0.37 \cite{KimSturzer13} or 0.81 \cite{NiStraszheim13}, i.e. significantly below the BCS-theory prediction of 1.43 in the weak-coupling limit. From the temperature dependencies of critical fields and the London penetration depth, the existence of multiple superconducting gaps has been suggested \cite{ChoTanatar12, WatsonMcCollam14, SurmachBrueckner14}, with possible indications that the gap anisotropy increases toward the edges of the superconducting dome as compared to the optimal doping \cite{ChoTanatar12}. The most recent $\mu$SR and NMR measurements \cite{SurmachBrueckner14} provided quantitative estimates of the gap magnitudes in the Pt-doped ``10-3-8'' compound, with the larger gap characterized by the $2\Delta/k_{\rm B}T_{\rm c}$ ratio only slightly higher than the BCS limit, and a tiny second gap that is at least an order of magnitude smaller in energy. It has already been suggested that iron-platinum arsenides could share the spin-fluctuation-mediated pairing mechanism with other iron-based superconductors \cite{ZhouKoutroulakis13, WatsonMcCollam14}. However, results of the NMR experiments suggested the presence of a pseudogap above $T_{\rm c}$ with a characteristic onset temperature, $T^\ast$, which was linked to an anomaly in the INS data resembling a precursor resonant mode \cite{SurmachBrueckner14}. This led to a suggestion of possible preformed Cooper pairing in these systems. Moreover, a Hebel-Slichter peak was found in the same work at very low temperatures ($\sim$\,2\,K) due to the opening of the smaller energy gap, which is therefore suspected to have $s$-wave symmetry that is distinct from the larger gap and is not typical for iron pnictides.

\section{Magnetic excitations in non-superconducting parent compounds}

\begin{wrapfigure}[20]{r}{0.51\textwidth}\vspace{-20pt}
\noindent\includegraphics[width=0.51\textwidth]{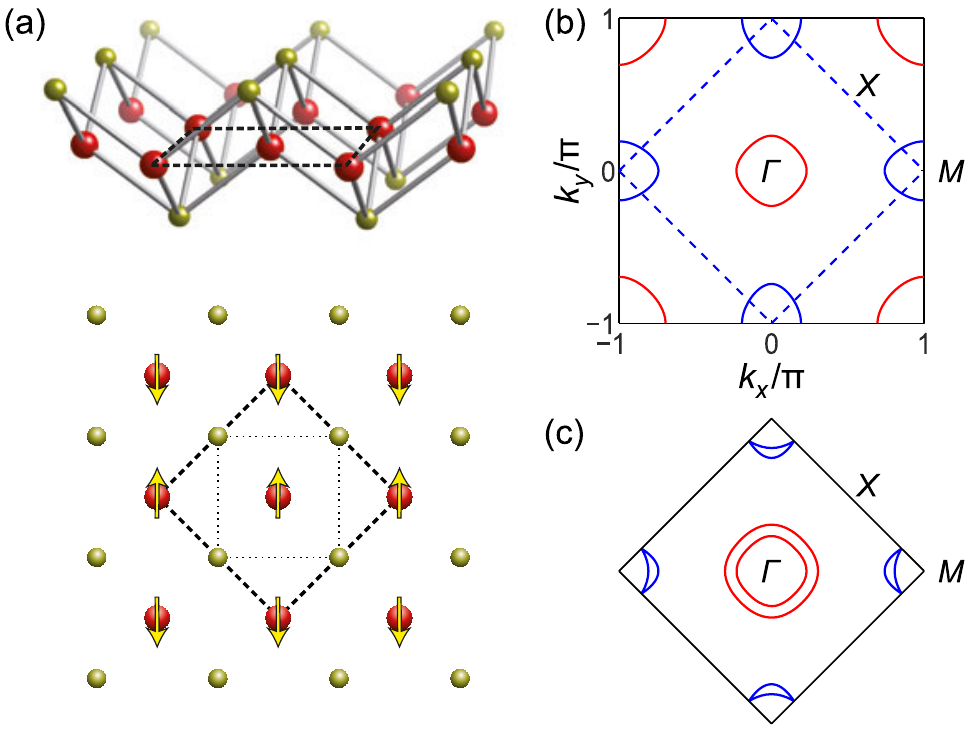}\vspace{-3.5pt}
\caption{(a)~Structure of the FeAs layer \cite{PaglioneGreene10}. Dashed and dotted lines denote the structural unit cells of the FeAs block of layers (two Fe atoms per unit cell) and of the Fe sublattice (one Fe atom per unit cell), respectively. (b, c)~Unfolded (top) and folded (bottom) Brillouin zones corresponding to one and two Fe atoms per unit cell, respectively \cite{RaghuQi08}.}
\label{Fig:FeAs-layer}
\end{wrapfigure}

\subsection{Folded vs. unfolded Brillouin-zone notations}

The unit cell of a typical iron-arsenide parent compound contains magnetic Fe atoms, which form a tetragonal sublattice at room temperature, and the remaining non-magnetic ions (As atoms and spacer layers) with generally lower lattice symmetry. For an individual FeAs block of layers, two formula units are contained in the unit cell because of the alternating positions of As atoms above and below the Fe plane, as illustrated in Fig.~\ref{Fig:FeAs-layer}\,(a). However, the unit cell of the simple square Fe sublattice is twice smaller, containing just one Fe atom per unit cell. As a result, the magnetic Fe sublattice possesses a higher symmetry in direct space with respect to the crystal itself. This justifies the introduction of a so-called unfolded Brillouin zone of the Fe sublattice in reciprocal space \cite{RaghuQi08} with a twice larger volume in comparison to the conventional crystallographic Brillouin zone, as shown in Fig.~\ref{Fig:FeAs-layer}\,(b,\,c).

Since electrons forming the conduction band are predominantly of Fe\,$d$ character \cite{KoitzschInosov08}, with both charge and magnetization density localized mainly around Fe atoms \cite{LeeVaknin10, BrownChatterji10}, it is generally expected that all essential physical quantities measured in the reciprocal space would follow the symmetry of the unfolded Brillouin zone. This has been demonstrated experimentally for the spin-excitation spectrum of ``122'' ferropnictides, where folding implies an intricate change of symmetry in three dimensions due to the body-centered tetragonal crystal structure peculiar to these systems~\cite{ParkInosov10}. Moreover, the folded Brillouin zone is insufficient for a proper description of magnetic structures or spin fluctuations, as it equates reciprocal-space points that may correspond to qualitatively different types of spin textures in direct space \cite{Johnston10}. For example, checkerboard AFM order with the $(\pi,\pi)$ propagation vector in the unfolded notation and simple ferromagnetic order residing at the zone center would both appear at the $\Gamma$ point of the folded zone. Therefore, for the sake of clarity, all reciprocal-space vectors in the present review will be given in the unfolded notation, unless explicitly stated otherwise, and expressed in reciprocal lattice units (r.l.u.) of the iron sublattice.

\subsection{Spin-wave spectrum and normal-state fluctuations}

\begin{table*}[b!]\vspace{-3pt}
\begin{minipage}{0.62\textwidth}\footnotesize
\begin{tabular}{l@{\hspace{-1em}}r@{~~~}l@{~~~}l@{\hspace{-0.5em}}r}\toprule
Compound & $T_{\rm N}$ (K) & $\mu_{\rm Fe}/\mu_{\rm B}$ & $\Delta_{\mathbf{Q}_{\rm AFM}}$ (meV) & Ref.\\
\midrule
K$_2$Fe$_4$Se$_5$   & 553 & 3.2 (Ref.\,\citenum{XiaoNandi13})            & ~~8.7(3) (smaller gap)& [\citenum{XiaoNandi13}]\\
                    &     &                                              & ~~16.5(3) (larger gap)& [\citenum{XiaoNandi13}]\\
SrFe$_2$As$_2$      & 205 & 1.0 (Ref.\,\citenum{KanekoHoser08, ZhaoRatcliff08, LeeVaknin10})            & ~~6.5 & [\citenum{ZhaoYao08}]\\
CaFe$_2$As$_2$      & 172 & 0.8 (Ref.\,\citenum{GoldmanArgyriou08})         & ~~6.9(2) & [\citenum{McQueeneyDiallo08, DialloPratt10}]\\
BaFe$_2$As$_2$      & 137 & 0.9 (Refs.\,\citenum{HuangQiu08,WilsonYamani09}) & ~~7.7(2) & [\citenum{EwingsPerring08}]\\
                    &     &                                      & ~~9.8(4) & [\citenum{MatanMorinaga09}]\\
                    &     &                                      & 10.1 (out-of-plane) & [\citenum{QureshiSteffens12}]\\
                    &     &                                      & 16.4 (in-plane) & [\citenum{QureshiSteffens12}]\\
                    &     &                                      & 10(1) & [\citenum{ParkFriemel12}]\\
LaFeAsO             & 140 & 0.4 (Ref.\,\citenum{CruzHuang08})            & ~~6.5 & [\citenum{RamazanogluLamsal13}]\\
Na$_{1-\delta}$FeAs & 45  & 0.1 (Refs.\,\citenum{LiCruz09, WrightLancaster12})& 10(2) & [\citenum{ParkFriemel12}]\\
BaFe$_{1.96}$Ni$_{0.04}$As$_2$ & 91 &                            & $\!\!\!\sim$\,2 & [\citenum{HarrigerSchneidewind09}]\\
\bottomrule
\end{tabular}
\end{minipage}\hfill
\begin{minipage}{0.36\textwidth}\vspace{9.5em}
\caption{Comparison of the N\'{e}el temperatures ($T_{\rm N}$), values of the ordered magnetic moment ($\mu_{\rm Fe}$), and zone-center gap energies ($\Delta_{\mathbf{Q}_{\rm AFM}}$) in several iron-arsenide compounds, updated after Ref.\,\citenum{ParkFriemel12}.}\label{Tab:AnisotropyGaps}
\end{minipage}
\vspace{-5pt}
\end{table*}

Above the N\'{e}el temperature, the spin-fluctuation spectrum of a typical iron-pnictide antiferromagnet consists of two gapless branches of paramagnon excitations that are sharply peaked at the $(\piup,0)$ and $(0,\piup)$ wave vectors, which coincide with the nesting vectors of the normal-state Fermi surface \cite{RaghuQi08}. At each of these two vectors,
\begin{wrapfigure}[20]{r}{0.7\textwidth}\vspace{-1pt}
\noindent\hspace{2.5pt}\includegraphics[width=0.7\textwidth]{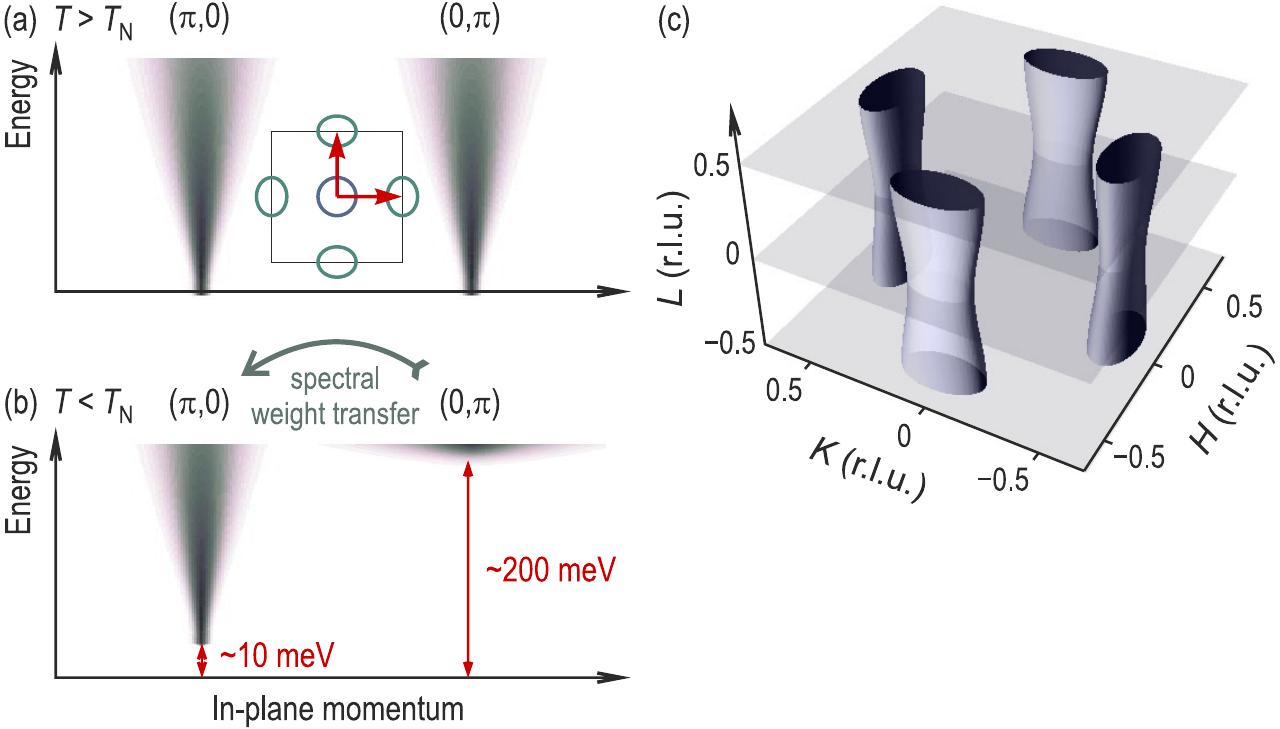}\vspace{-56pt}
\mbox{\hspace{0.38\textwidth}\begin{minipage}{0.32\textwidth}
\caption{(a,\hspace{1pt}b)~Schematic illustration of the observable changes in the spin-fluctuation spectrum of a typical iron-pnictide antiferromagnet upon crossing the N\'{e}el temperature. (c)~The shape of paramagnon excitations in the three-dimensional $\mathbf{Q}$ space, schematically illustrated by warped cylinders representing constant-intensity contours of the INS signal.}\label{Fig:GapsSketch}
\end{minipage}}
\end{wrapfigure}
low-energy excitations disperse very steeply and therefore can be seen in a thermal-neutron scattering experiment as a single commensurate peak that broadens towards higher energies, as it is schematically illustrated in Fig.~\ref{Fig:GapsSketch}\,(a). This peak usually exhibits an elliptical cross-section within the $Q_x\,Q_y$ plane in momentum space, whose aspect ratio reflects the nesting properties of the normal-state Fermi surface and therefore directly depends on the electron doping level of the material \cite{ParkInosov10}. In the undoped ``122'' compounds, it is typically elongated in the transverse direction with respect to the momentum vector \cite{ParkInosov10, DialloPratt10, HarrigerLuo11, HarrigerLiu12} with the only exception of SrCo$_2$As$_2$, where it extends longitudinally like in Ba$_{1-x}$K$_x$Fe$_2$As$_2$ \cite{ZhangWang11} despite the expected electron-doped character of the Fermi surface \cite{JayasekaraLee13}. Along the $Q_z$ direction, the peak shape remains unchanged \cite{ParkInosov10}, whereas its intensity oscillates, reflecting the three-dimensional character of magnetic exchange interactions. In BaFe$_2$As$_2$ or CaFe$_2$As$_2$, for example, the intensity maxima are located at the $(\half\,0\,L)$ and $(0\,\half\,L)$ wave vectors with half-integer $L$ components \cite{DialloPratt10, ParkFriemel12}, coinciding with the low-temperature ordering vectors. Still, the intensity remains finite (only 2--3 times weaker) even at integer $L$ values. As a result, in three dimensions the paramagnon excitations can be imagined as corrugated rods of intensity with an elliptical cross-section [see Fig.\,\ref{Fig:GapsSketch}\,(c)]. In the itinerant approach, this corrugation results from a variation of nesting properties along $Q_z$ as a consequence of Fermi surface warping in the $k_z$ direction, and is qualitatively well captured in density-functional-theory calculations \cite{ParkInosov10}.

Upon crossing $T_{\rm N}$ into the stripe-AFM ordered phase, the magnetic excitation spectrum undergoes two characteristic changes illustrated in Fig.~\ref{Fig:GapsSketch}\,(b). First, a small anisotropy gap opens up on a characteristic energy scale of the order of $10$\,meV near the magnetic zone center \cite{McQueeneyDiallo08, DialloAntropov09, QureshiSteffens12}. This gap originates from the
\begin{wrapfigure}[31]{r}{0.37\textwidth}\vspace{-4.5pt}
\noindent\includegraphics[width=0.37\textwidth]{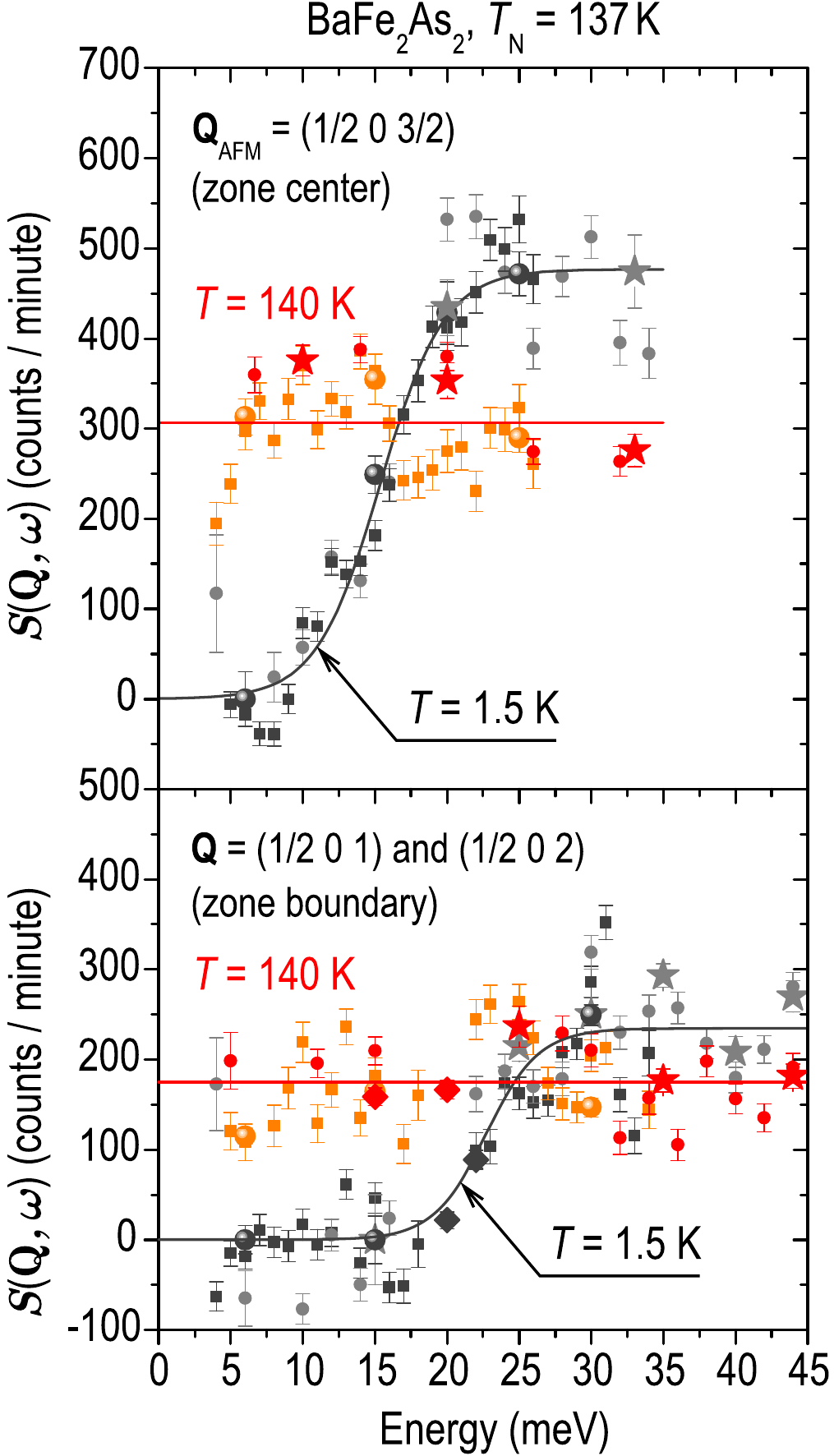}\vspace{-3pt}
\caption{Energy dependence of the scattering function, $S(\mathbf{Q},\omega)$, measured for BaFe$_2$As$_2$ above and below $T_N$ at the ordering wave vector ($L = \threehalf$) and at the magnetic zone boundary ($L = 1, 2$) \cite{ParkFriemel12}.}
\label{Fig:GapDispersion}
\end{wrapfigure}
broken Heisenberg symmetry of the magnetic moments induced by exchange anisotropy and does not require any detwinning procedure to be detected. Remarkably, the magnitude of this gap does not vary much among different families of iron pnictides despite their very different values of $T_{\rm N}$ and the ordered magnetic moment, but can be rapidly suppressed by doping, as seen from Table~\ref{Tab:AnisotropyGaps}. For instance, in the insulating K$_2$Fe$_4$Se$_5$ with a large $T_{\rm N}=553$\,K and an ordered moment of $\sim$\,3.2$\mu_{\rm B}$ \cite{XiaoNandi13}, the anisotropy gap is nearly the same as in Na$_{1-\delta}$FeAs with $T_{\rm N}=45$\,K and its tiny ordered moment of only $\sim$\,0.1$\mu_{\rm B}$. This universality goes in contrast with the general expectation that the anisotropy gap should scale with the sublattice magnetization following a simple power law relationship \cite{FishmanLiu98}, and its explanation still represents a challenge for the theory. As a function of $Q_z$, the zone-center gap
undergoes only a rather weak increase up to $\sim$\,20\,meV at integer $L$ values (magnetic zone boundary in the $z$ direction) \cite{ParkFriemel12, SongRegnault13}. In the terminology of the anisotropic Heisenberg model employed for an empirical description of the spin-wave spectrum \cite{HarrigerLuo11}, this indicates that the zone-boundary gap represents a combined effect of the out-of-plane exchange coupling, $J_{\rm c}$, and the single-ion anisotropy constant, $J_{\rm s}$, that are comparable in magnitude \cite{ParkFriemel12}. Consequently, the anisotropy term should not be neglected in any realistic description of the low-energy magnon spectrum in three dimensions. It is even more remarkable that the variation of the spin gap in the $Q_z$ direction anticorrelates with the above-mentioned modulation of intensity, while both quantities change by about a factor of two between integer and half-integer $L$ in BaFe$_2$As$_2$. As a result, the amount of spectral weight transferred from the spin-gap region upon crossing $T_{\rm N}$ is almost independent of $Q_z$, as can be seen in Fig.\,\ref{Fig:GapDispersion}. Whether this observation, which emphasizes the highly two-dimensional nature of magnetism in ferropnictides, is accidental or holds for a broader class of compounds still remains to be verified.

The second important transformation that should happen with the magnon spectrum upon crossing the N{\'e}el temperature is associated with the broken fourfold rotational symmetry peculiar to the stripe-AFM magnetic order. In the AFM phase, one of the two (initially equivalent) wave vectors, $(\piup,0)$ or $(0,\piup)$, is spontaneously selected by the system as the magnetic propagation vector. Let us assume for the sake of definiteness that this wave vector is $(\piup,0)$, which puts the remaining $(0,\piup)$ wave vector at the magnetic zone boundary. According to the localized Heisenberg-type description of the spin-wave spectrum \cite{ZhaoAdroja09, HarrigerLuo11}, the magnon energy at this point reaches its maximal value of about 200\,meV. Therefore, this model suggests that the whole spectral weight of the $(0,\piup)$ branch of paramagnon excitations (up to $\sim$200\,meV), which was initially gapless above $T_{\rm N}$, is fully transferred to the $(\piup,0)$ wave vector in the ordered state, where the intensity should double accordingly. However, in a real experiment this dramatic effect is usually masked by the presence of magnetic domains with both orientations in the sample, unless special detwinning procedures are used \cite{LuPark14}. Until now, it has not been possible to demonstrate the vanishing of the zone-boundary spectral weight in the whole energy range, and the detailed temperature dependence of the $(0,\piup)$ spectrum also remains unclear.\enlargethispage{6pt}

\begin{wrapfigure}[20]{r}{0.55\textwidth}\vspace{-3pt}
\noindent\includegraphics[width=0.55\textwidth]{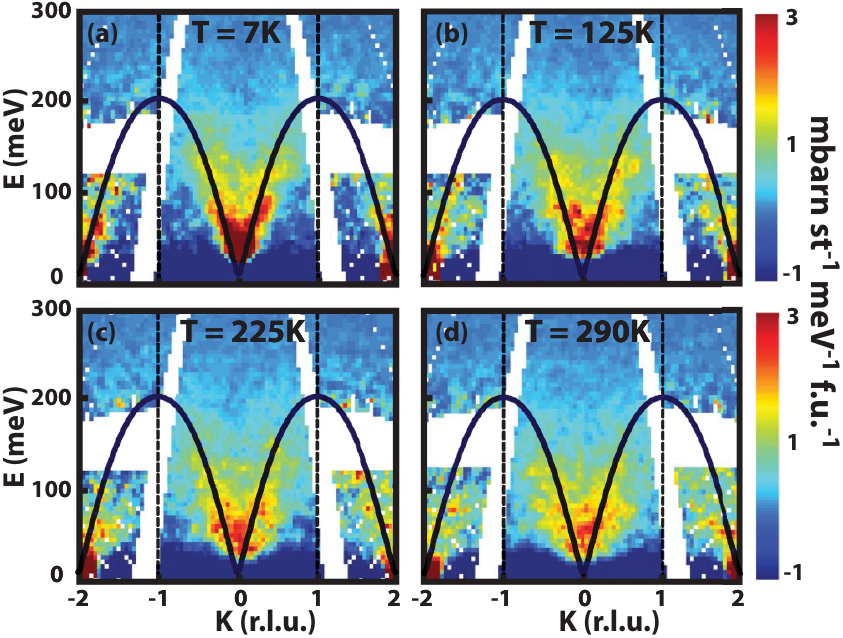}\vspace{-3.5pt}
\caption{Temperature dependence of the magnetic excitation spectrum in BaFe$_2$As$_2$. The solid line is the spin-wave dispersion fitted to a Heisenberg model. Reproduced from Ref.~\citenum{HarrigerLiu12}, copyright by the American Physical Society.}
\label{Fig:BFA-Tdep}
\end{wrapfigure}

In a magnetically twinned sample, the experimental spin-excitation spectrum evolves rather gradually across $T_{\rm N}$, experiencing no evident changes above the anisotropy gap apart from a monotonic broadening of the signal with temperature \cite{EwingsPerring11, HarrigerLiu12}, as shown in Fig.\,\ref{Fig:BFA-Tdep} for the BaFe$_2$As$_2$ parent compound. In energy, one can arbitrarily define three characteristic regions: (i) low-energy region below $\sim$\,50\,meV, where the two spin-wave branches 
near the AFM wave vector can not be resolved, resulting in a single commensurate peak in a constant-energy scan \cite{HarrigerLiu12}; (ii) intermediate-energy region between $\sim$\,50\,meV and $\sim$\,150\,meV, where two broad dispersive peaks can be clearly observed, and (iii) high-energy region near the top of the spin-wave branch, which is strongly broadened by interactions with the particle-hole continuum. While a very broad peak can be still recognized around 150--200\,meV in energy near the zone boundary \cite{HarrigerLuo11, HarrigerLiu12}, the magnetic signal at these energies is very much overdamped and devoid of any clear momentum-space structure. This strong damping of the zone-boundary magnons can be qualitatively captured by theories that either assume a strong coupling of local moments in the Heisenberg model to itinerant electrons \cite{LeongLee14} or describe the spin waves from an itinerant perspective as collective excitations of the SDW state treated within the random phase approximation (RPA) \cite{KnolleEremin10, RoweKnolle12}. Which one of the two approaches presents a generally better description of the spin-wave spectrum in undoped ferropnictides still remains a subject of debate, yet it is clear that the consideration of itinerant electrons is important even in local-moment models \cite{LeongLee14}.

Finally, it is worth mentioning that the magnon spectrum has been also considered theoretically for an alternative orthomagnetic ordering model of iron pnictides that resides at the same $(\piup,0)$ and $(0,\piup)$ wave vectors, but causes no orthorhombic distortion of the lattice \cite{WangKang15}. This model was initially motivated by the experimental observations of tetragonal antiferromagnetic phases in Ba(Fe$_{1-x}$Mn$_x$)$_2$As$_2$ \cite{KimKreyssig10} and Ba$_{1-x}$Na$_x$Fe$_2$As$_2$ \cite{AvciChmaissem14}.

\subsection{Spin anisotropy of magnon excitations}\enlargethispage{6pt}

The magnetic neutron-scattering intensity that contributes to the total signal in a non-spin-polarized INS experiment can be additionally decomposed into several terms that differ by the direction of spin polarization and can be distinguished by means of the neutron polarization analysis: out-of-plane, in-plane transverse, and in-plane longitudinal with respect to the orientation of the magnetic moments. Spin anisotropy refers to the difference between the scattering functions in orthogonal spin-polarization channels, which originates from the spin-orbit interaction that breaks the isotropic Heisenberg symmetry in the spin space. For layered materials, such
\begin{wrapfigure}[18]{l}{0.51\textwidth}\vspace{-2pt}
\noindent\includegraphics[width=0.51\textwidth]{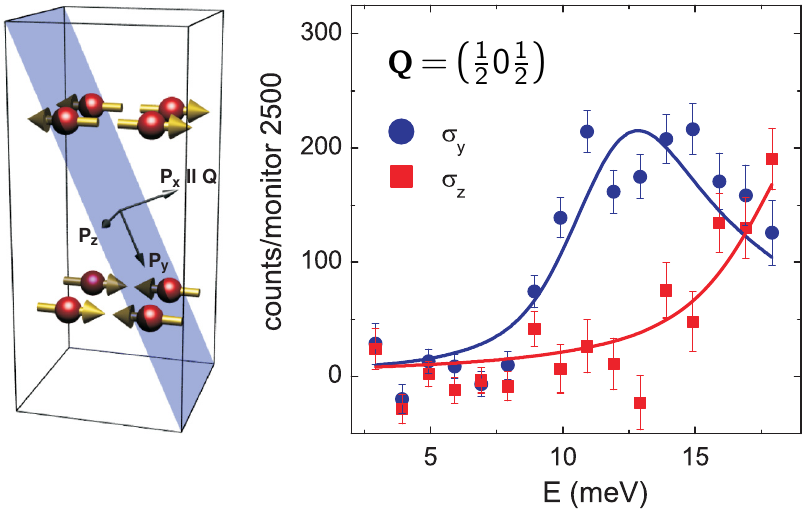}\vspace{-10pt}
\caption{Definition of the reference frame for the neutron polarization analysis on the $(\half 0 \half)$ magnetic wave vector (left panel) and the corresponding scattering cross-sections with $\sigma_y$ and $\sigma_z$ polarizations in BaFe$_2$As$_2$, resulting from the neutron polarization analysis (right panel). Adapted from Ref.\,\citenum{QureshiSteffens12}, copyright by the American Physical Society.}
\label{Fig:BFA-GapAnisotropy}
\end{wrapfigure}
as copper-oxide parent compounds of high-$T_{\rm c}$ superconductors or iron pnictides, spin anisotropy is represented by the difference between magnetic fluctuations with in-plane and out-of-plane polarizations. Yet, in contrast to undoped cuprates, which are typically characterized by an easy-plane anisotropy \cite{PetersBirgeneau88, TranquadaShirane89, ShamotoSato93, RossatMignot93, BourgesSidis94, PetitgrandMaleyev99}, in iron pnictides it costs less energy to tilt the spins out of the FeAs plane than to rotate them within the $ab$-plane \cite{QureshiSteffens12}. Consequently, the anisotropy gap in pnictides is larger for the in-plane polarization \cite{QureshiSteffens12, SongRegnault13}, resulting in the lowest-energy magnon excitations being fully $c$-axis polarized below $T_{\rm N}$ (see Fig.\,\ref{Fig:BFA-GapAnisotropy}). First-principles calculations are still faced with apparent difficulties in reproducing these different energy scales in the spin-wave spectra even on a qualitative level \cite{YareskoLiu09, QureshiSteffens12}, unless the magnitudes of the ordered magnetic moments are artificially scaled down.

\begin{wrapfigure}[32]{r}{0.41\textwidth}\vspace{-2pt}
\noindent\includegraphics[width=0.41\textwidth]{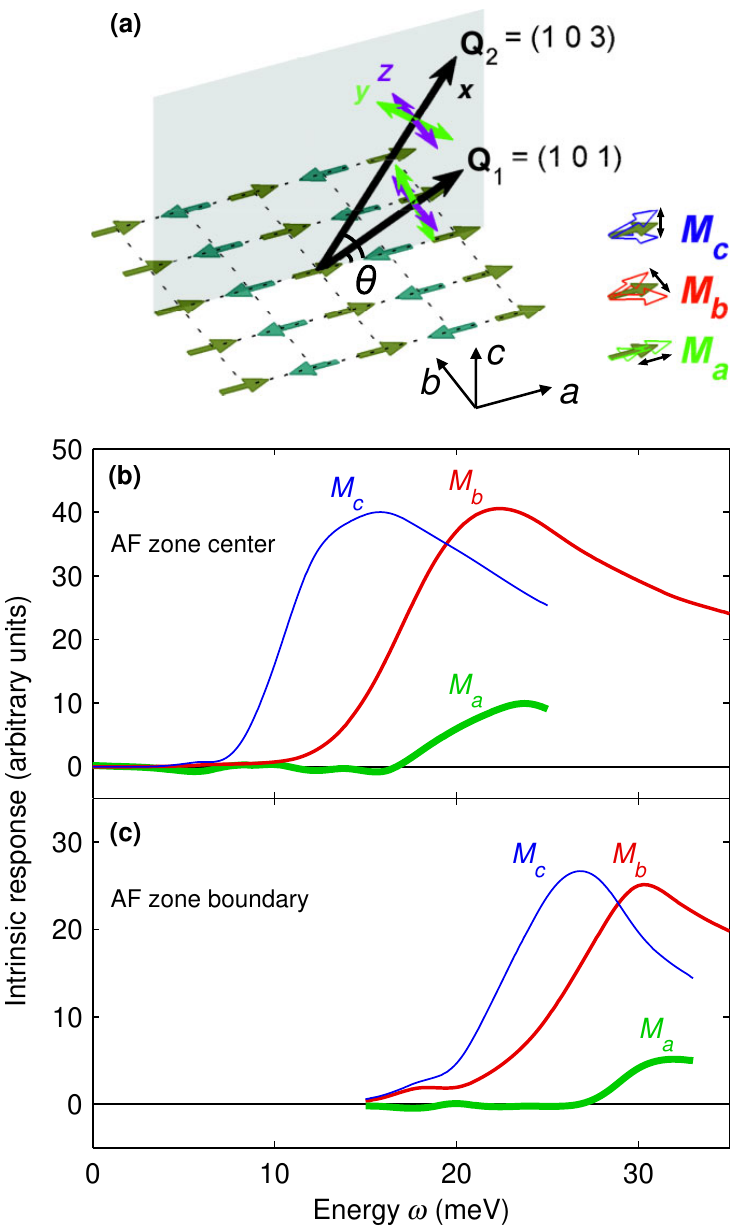}\vspace{-5pt}
\caption{(a)~Spin arrangement in the AFM phase of BaFe$_2$As$_2$ and possible fluctuation directions. (b,c)~Interpolated magnetic responses for three different polarizations that contribute to the scattering function at the magnetic zone center and at the zone boundary. Adapted from Ref.\,\citenum{WangZhang13prx}, copyright by the American Physical Society.}
\label{Fig:Longitudinal}
\end{wrapfigure}\enlargethispage{5pt}
From experiments on the NaFeAs parent compound, in which the structural and magnetic phase transitions are decoupled, it has been suggested that the low-energy spin excitations become anisotropic within the $ab$ plane for $T_{\rm N} < T < T_{\rm s}$, restoring the isotropic behavior only above the structural transition temperature \cite{SongRegnault13}. Nevertheless, our own studies on the same compound indicated that the anisotropy gap in the spin-wave spectrum closes at $T_{\rm N}$, so that the spectrum remains gapless in the nematic phase \cite{ParkFriemel12}. In slightly electron-underdoped BaFe$_{1.904}$Ni$_{0.096}$As$_2$, the in-plane spin anisotropy of paramagnon excitations could be observed even above the structural transition temperature in the absence of a uniaxial pressure \cite{LuoWang13}. Polarized neutron-spectroscopy experiments have additionally indicated the presence of longitudinal spin fluctuations with the polarization direction parallel to the ordered moments \cite{WangZhang13prx, SongRegnault13}. The presence of such excitations is disallowed for spin waves in a classical local-moment Heisenberg model, and was therefore attributed to itinerant magnetism of the SDW type \cite{WangZhang13prx}. The longitudinal component is characterized by the largest spin gap of about 18\,meV at the zone center (see Fig.\,\ref{Fig:Longitudinal}). The same work also reports the dispersion of the spin gap along the $Q_z$ direction for different spin channels. As can be seen from comparing Figs.\,\ref{Fig:Longitudinal}\,(b) and (c), the difference in energy of about 10\,meV between the spin gaps at integer and half-integer $L$ remains approximately the same for \mbox{all three polarization directions}.

\subsection{Effects of impurities}

In moderately electron-doped ``122'' compounds that still preserve the AFM order with a reduced $T_{\rm N}$, initially well-defined spin waves evolve with an increasing dopant concentration towards more overdamped and diffusive excitations, this crossover being approximately concomitant with the onset of superconductivity \cite{TuckerFernandes14}. The observable changes in the spectrum are manifested by a rapid reduction and smearing of the anisotropy gap, a change of the spectral shape towards a more broadened diffusive response \cite{TuckerFernandes14}, and reduced three-dimensionality \cite{HarrigerSchneidewind09}. For example, at only 4\% Ni doping, the spin gap in BaFe$_2$As$_2$ experiences a nearly fivefold reduction to $\sim$\,2\,meV despite a much more moderate suppression of $T_{\rm N}$ by only 50\% from initial 137\,K to 91\,K \cite{HarrigerSchneidewind09}. Still, the magnetic spectral weight remains partially depleted in the whole low-energy region up to the initial gap energy ($\sim$\,10\,meV), which can be thought of as a magnetic pseudogap resulting from an inhomogeneous distribution of the local anisotropy gaps within the sample. Ultimately, at higher doping levels, this inhomogeneity develops into a cluster spin glass phase in the coexistence region of the phase diagram, characterized by a smeared N\'{e}el transition \cite{LuTam14}. In this regime, short-range incommensurate AFM ordered regions were shown to coexist on a mesoscopic scale and compete with superconductivity, as has been demonstrated recently for nearly optimally doped BaFe$_{2-x}$Ni$_x$As$_2$ \cite{LuTam14}. This situation is reminiscent of the phase-separation phenomena in hole-doped ``122'' pnictides, discussed shortly after the discovery of iron-based superconductivity \cite{ParkInosov09}. In the vicinity of $T_{\rm N}$, high-resolution neutron resonance spin-echo (NRSE) spectroscopy has further revealed an increase in the energy width of the $(\half 0 \threehalf)$ magnetic Bragg reflection, evidencing slow fluctuations of the magnetic order parameters, similar to the dynamics of a typical spin glass \cite{LuTam14}. Consequently, the different characteristic time scales of various employed probes, such as Mössbauer spectroscopy, NRSE, elastic triple-axis neutron scattering, or conventional neutron diffraction, led to a noticeable mismatch in the experimentally determined spin-freezing temperatures and differing shapes of the magnetic transition. Moreover, the spatial correlation length of the magnetic order parameter evolved rather smoothly and did not diverge down to the lowest measured temperatures, \mbox{showing no indications of a sharp phase transition \cite{LuTam14}}.

\begin{figure*}[t!]\vspace{-4pt}
\begin{minipage}[b]{0.65\textwidth}
\includegraphics[width=\textwidth]{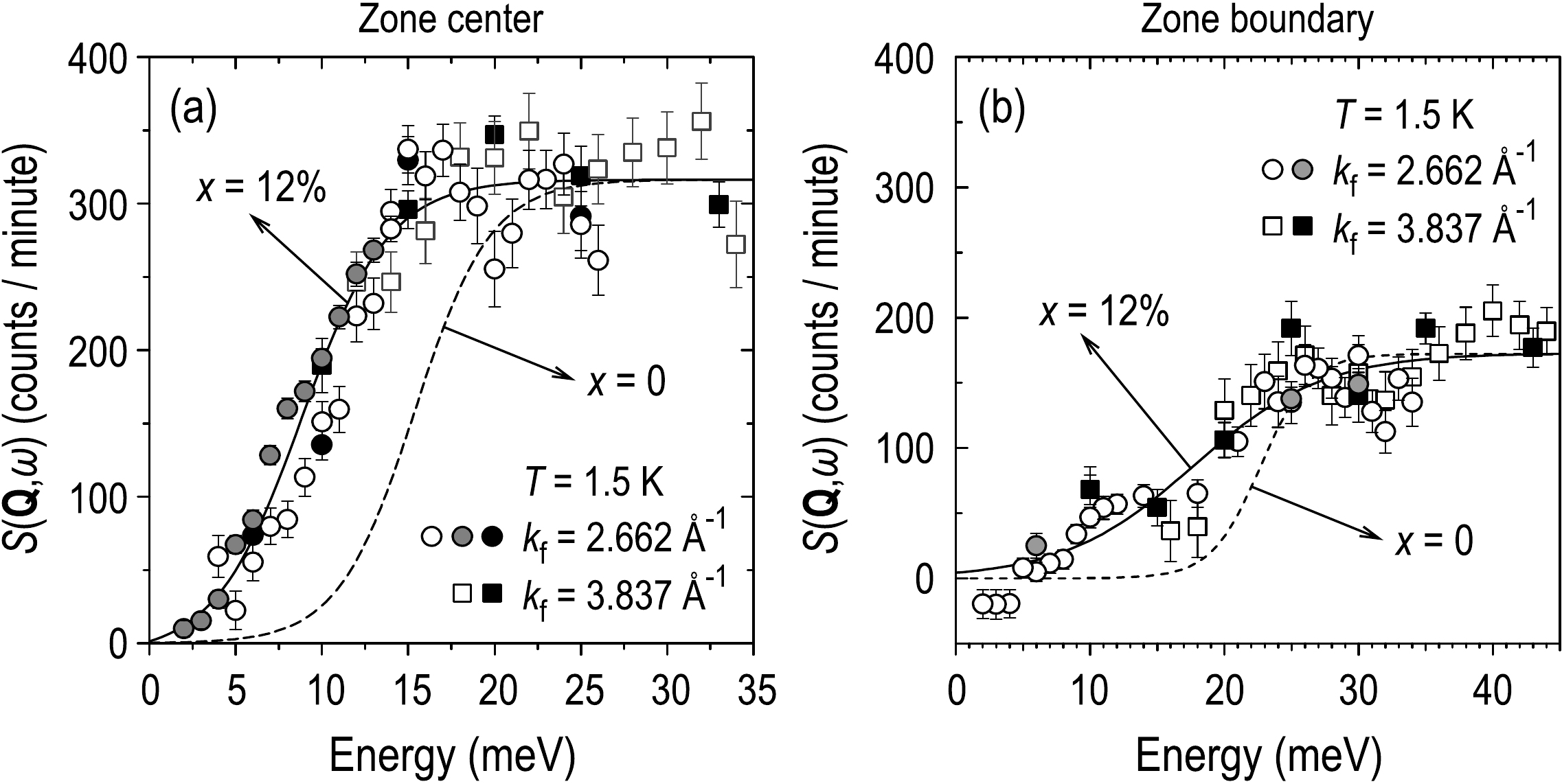}
\end{minipage}\hfill
\begin{minipage}[b]{0.32\textwidth}
\caption{\footnotesize Reduction of the anisotropy gap in BaFe$_2$As$_2$ by Mn doping at the magnetic zone center (half-integer $L$, left panel) and at the zone boundary (integer $L$, right panel). The corresponding spectral function of the parent compound is plotted with dashed lines for comparison. Reproduced from Ref.\,\citenum{InosovFriemel13}, copyright by the American Physical Society.\vspace{9em}}
\label{Fig:BFMA-gaps}
\end{minipage}\vspace{-10pt}
\end{figure*}

A similarly complex behavior is observed with the introduction of magnetic impurities, as illustrated by the most studied case of Ba(Fe$_{1-x}$Mn$_x$)$_2$As$_2$, in which Mn ions carry a large local magnetic moment \cite{TexierLaplace12}. While Mn doping is detrimental to superconductivity, it causes notable changes in the spin-wave spectrum of BaFe$_2$As$_2$. Below a certain critical Mn concentration of $x_{\rm c}\approx10$\%, the AFM phase transition remains sharp and is gradually suppressed with the dopant concentration, as for many other dopant atoms \cite{KimKreyssig10, KimPratt11}. Above $x_{\rm c}$, the transition abruptly smears out with the formation of a cluster spin glass phase that nucleates at temperatures far above the initial $T_{\rm N}$ of the parent compound. Long-range order sets in at lower temperatures, where it continues to coexist with the paramagnetic regions over a broad temperature range, resulting in Griffiths-type behavior \cite{InosovFriemel13}. Concurrently with the phase transition smearing, we observed a considerable reduction and broadening of the anisotropy gap, as illustrated in Fig.\,\ref{Fig:BFMA-gaps}. Similarly to the situation with Ni doping, the onset of the spin gap is reduced nearly to zero in 12\% Mn-doped sample both for zone-center and zone-boundary magnons. However, a partial reduction of the magnetic spectral weight (spin pseudogap) is still observed over the whole low-energy range comparable with the gap magnitude in the parent compound. The smearing of the gap onset appears to be stronger near the zone boundary, where the gap was initially larger [see Fig.\,\ref{Fig:BFMA-gaps}\,(b)].

\begin{figure*}[b!]\vspace{-10pt}
\begin{minipage}[b]{0.73\textwidth}\hspace{-4pt}
\hspace{-3pt}\includegraphics[width=\textwidth]{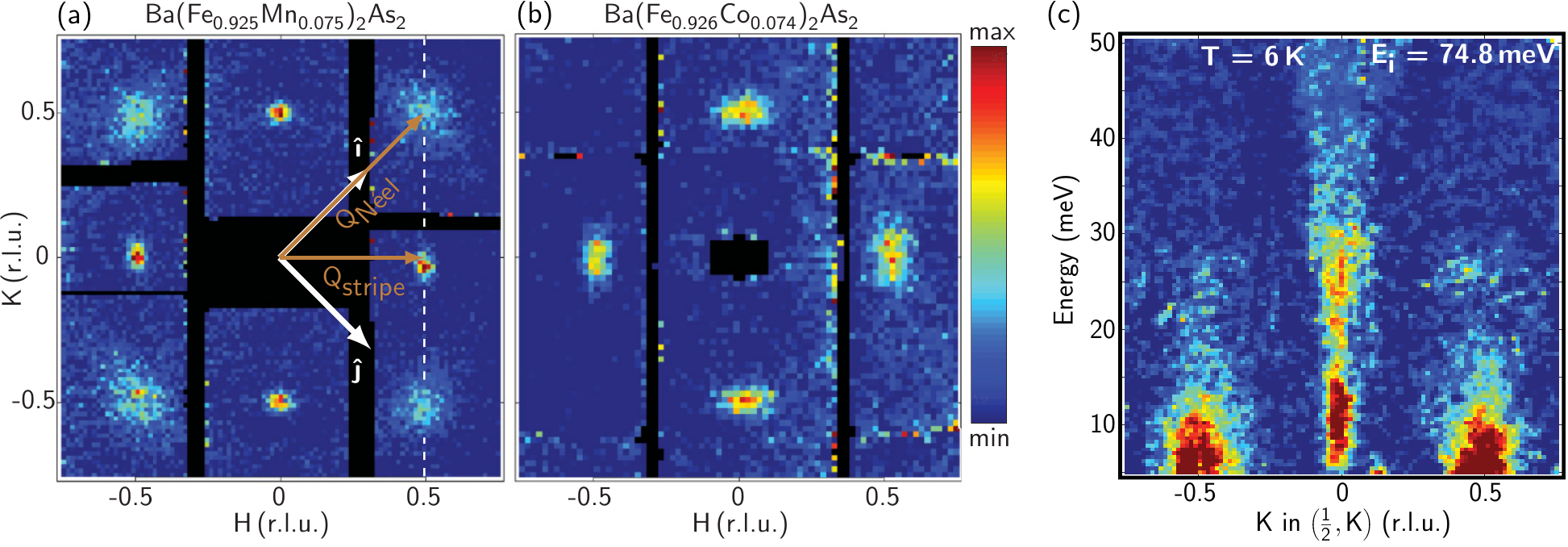}
\end{minipage}\hfill
\begin{minipage}[b]{0.255\textwidth}
\caption{\footnotesize (a,b)~Diffuse scattering intensity at the $\left(\kern-.5pt\frac{1}{2}\frac{1}{2}\kern-.5pt\right)$ wave vector in Ba(Fe$_{0.925}$Mn$_{0.075}$)$_2$As$_2$, evidencing N{\'e}el-type fluctuations induced by the magnetic Mn impurities, as opposed to Ba(Fe$_{0.926}$Co$_{0.074}$)$_2$As$_2$. (c)~The energy spectrum of the $\left(\kern-.5pt\frac{1}{2}\frac{1}{2}\kern-.5pt\right)$ and $\left(\kern-.5pt\frac{1}{2}0\kern-.5pt\right)$ spin fluctuations, plotted along the dashed line in panel (a). Reproduced from Ref.\,\citenum{TuckerPratt12}, copyright by the American Physical Society.\vspace{1em}}
\label{Fig:BFMA-PiPi}
\end{minipage}
\end{figure*}

A recent work performed on a 7.5\% Mn-doped sample ($x<x_{\rm c}$, $T_{\rm N}=80$\,K) has also indicated that introduction of Mn local moments could nucleate fluctuating short-range AFM order with a N{\'e}el-type structure at $(\piup,\piup)$ as opposed to the conventional stripe-like $(\pi, 0)$ order of the parent compound \cite{TuckerPratt12}. This is evidenced by the appearance of an additional diffuse inelastic scattering intensity around the $(\half\half)$ wave vector, as illustrated in Fig.\,\ref{Fig:BFMA-PiPi}. Due to the kinematic constraints of the time-of-flight (TOF) experiment with a fixed sample position, employed by G.\,S.\,Tucker \textit{et al.} \cite{TuckerPratt12}, the energy transfer is coupled to the out-of-plane component of the momentum, $L$. This results in a considerable variation of $L$ with energy in Fig.\,\ref{Fig:BFMA-PiPi}\,(c), which distorts the energy shape of the spectrum as compared to a constant-$\mathbf{Q}$ measurement. Therefore, the exact energy dependence of the $(\half\half)$ magnetic intensity, as well as its behavior at low energies within the spin-gap region below $5$\,meV still remains unresolved and needs to be investigated.\enlargethispage{6pt}

The effects of both nonmagnetic and magnetic impurities on the spin structure of iron pnictides has been also investigated theoretically. For instance, is was shown using Monte Carlo simulations that the introduction of nonmagnetic impurity sites into the Fe sublattice can lead to the formation of anticollinear magnetic order, i.e., it can qualitatively alter the magnetic ground state of the material \cite{WeberMila12}. It has been also demonstrated that individual magnetic impurities can exhibit cooperative behavior due to the Ruderman-Kittel-Kasuya-Yosida interaction mediated by conduction electrons, which in the case of Mn-doped BaFe$_2$As$_2$ leads to the short-range checkerboard $(\piup,\piup)$ spin fluctuations observed in experiments along with the more conventional stripe-like fluctuations \cite{GastiasoroAndersen14}. Similar results were also obtained for nonmagnetic impurities doped into the superconducting LiFeAs \cite{GastiasoroHirschfeld13}, where they may produce sub-gap bound states or locally pin the orbital and magnetic order to nucleate static magnetic clusters that are not present in the stoichiometric LiFeAs. As follows from the entire corpus of accumulated experimental and theoretical evidence, introduction of impurities offers a convenient ``tuning knob'' to shift the balance between different competing ground states and to explore qualitatively new magnetic states in the phase diagram of layered iron pnictides.\vspace{-2pt}\enlargethispage{5pt}

\section{Spin dynamics in superconducting compounds above and below $T_{\rm c}$}

\subsection{Neutron spin resonance and normal-state fluctuations}

\begin{figure*}[t!]\vspace{-1pt}
\includegraphics[width=\textwidth]{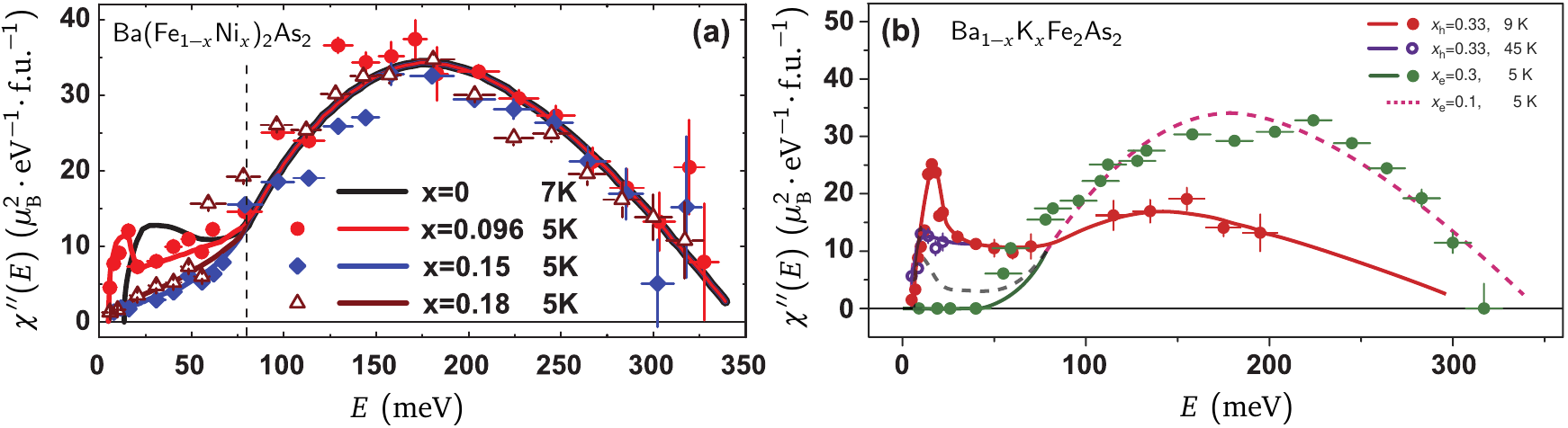}\vspace{-2pt}
\caption{Energy dependence of the total momentum-integrated magnetic spectral weight in electron- and hole-doped ``122'' compounds, presented in absolute units. Adapted from Ref.~\citenum{LuoLu13} (copyright by the American Physical Society) and Ref.~\citenum{WangZhang13}.\vspace{-3pt}}
\label{Fig:Q-integrated}
\end{figure*}

Next, we consider spin excitations in doped nonmagnetic iron-based superconductors above and below $T_{\rm c}$. It has been established from both high-energy neutron TOF spectroscopy \cite{LuoLu13, WangZhang13} and resonant inelastic X-ray scattering (RIXS) \cite{ZhouHuang13} that strongly overdamped paramagnon branches resembling the spin waves of the parent compounds persist over the entire Brillouin zone of iron pnictides near the optimal doping and beyond. This situation resembles the well-known case of high-$T_{\rm c}$ cuprates, where damped dispersive spin-wave-like modes 
\begin{wrapfigure}[20]{r}{0.5\textwidth}\vspace{-2pt}
\noindent\includegraphics[width=0.5\textwidth]{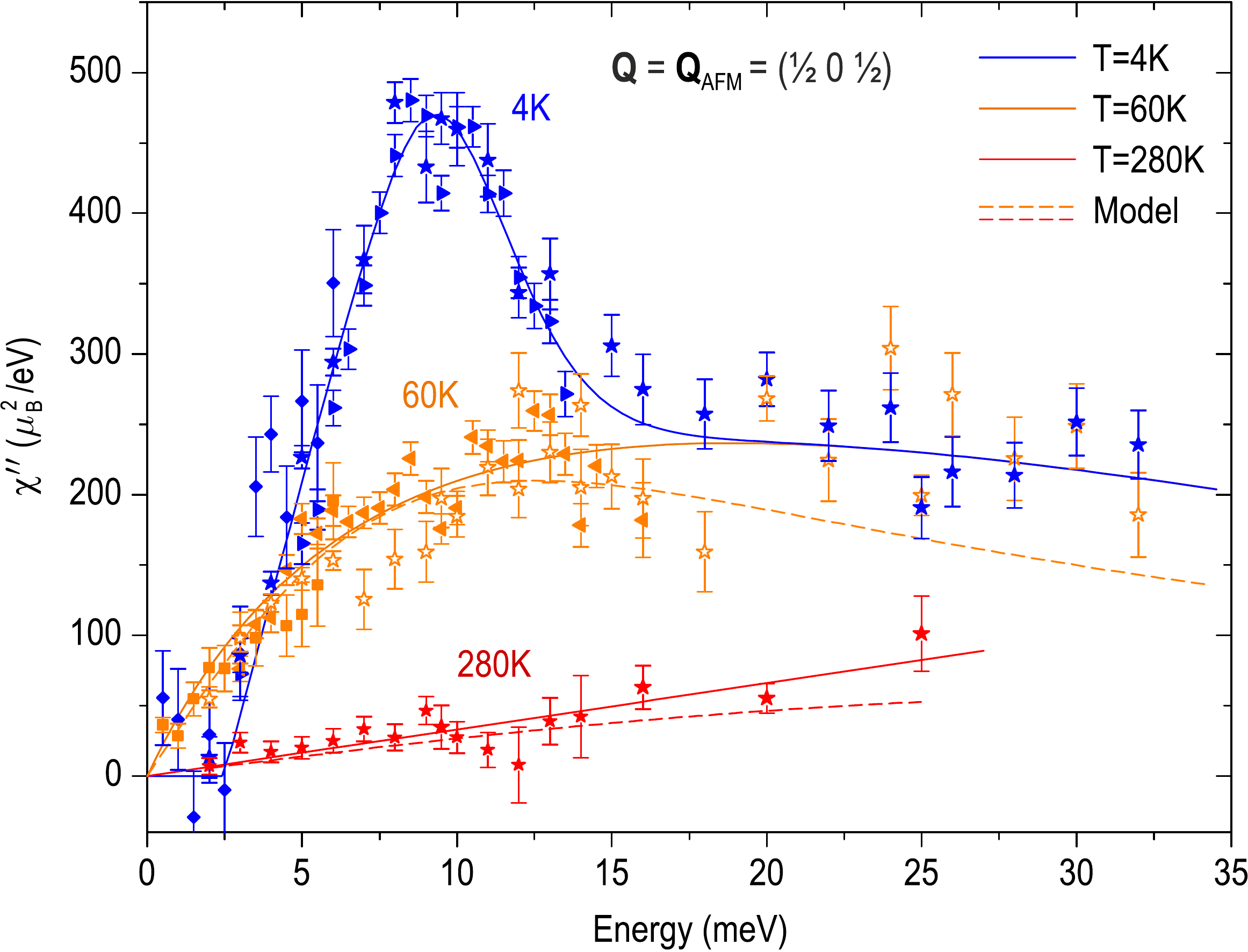}\vspace{-5pt}
\caption{Imaginary part of the spin susceptibility in the superconducting ($T=4$\,K) and the normal state ($T=60$ and 280\,K) of BaFe$_{1.85}$Co$_{0.15}$As$_2$ ($T_{\rm c}=25$\,K), obtained after the Bose-factor correction of the scattering function. Reproduced from Ref.\,\citenum{InosovPark10}.}
\label{Fig:Resonance}
\end{wrapfigure}
were found by RIXS in under- and overdoped compounds with essentially unchanged integrated spectral weight as compared to the parent antiferromagnetic insulators \cite{LeTaconGhiringhelli11}. A vanishingly weak effect of electron doping on spin fluctuations was also suggested from the results of a recent $^{75}$As NMR study \cite{GrafeGraefe14}, in which the pure BaFe$_2$As$_2$ parent compound was compared with doped Ba(Fe$_{1-x}$Co$_{x}$)$_2$As$_2$ and Ba(Fe$_{1-x}$Cu$_{x}$)$_2$As$_2$ in terms of the temperature dependence of their nuclear spin-lattice relaxation rate, $(T_1T)^{-1}$. Indeed, while high-energy paramagnons could be to a certain extent relevant for the superconducting pairing \cite{LeTaconGhiringhelli11}, in electron-doped ``122'' iron pnictides the total $\mathbf{Q}$-integrated magnetic spectral weight above $\sim$\,80\,meV in energy remains surprisingly insensitive to doping \cite{LuoLu13}, as shown in Fig.\,\ref{Fig:Q-integrated}\,(a), unlike in the hole-doped Ba$_{1-x}$K$_x$Fe$_2$As$_2$ \cite{WangZhang13}, where it is considerably suppressed [see Fig.\,\ref{Fig:Q-integrated}\,(b)]. Still, the most pronounced changes occur mainly in the low-energy region of the spectrum with either doping or temperature. As can be seen in Fig.\,\ref{Fig:Q-integrated}, both electron and hole doping leads to an overall reduction of the spectral weight below $\sim$\,80\,meV. However, in the hope-doped Ba$_{1-x}$K$_x$Fe$_2$As$_2$ the low-energy spectral weight remains finite at all doping levels including the end-member KFe$_2$As$_2$, while in the strongly electron-overdoped nonsuperconducting BaFe$_{1.7}$Ni$_{0.3}$As$_2$ no magnetic spectral weight is present up to $\sim$\,50\,meV~\cite{WangZhang13}.

As already mentioned before, the most prominent signature of superconductivity in the spin-excitation spectrum of unconventional superconductors is the magnetic resonant mode, which can be seen in Fig.\,\ref{Fig:Q-integrated} as a sharp low-energy peak that is observed only in superconducting samples below $T_{\rm c}$. A closer look at the low-energy part of the spectrum (Fig.\,\ref{Fig:Resonance}) reveals a depletion of spectral weight below the resonance, leading to a spin gap opening and a consequent transfer of the spectral weight to the resonance peak. The same figure also illustrates that the normal-state spectrum at low energies can be well described by the predictions of the theory of nearly antiferromagnetic metals (dashed lines) \cite{Moriya85}. The presence of the resonant mode has now been established for many structurally distinct families of iron-based superconductors, and for some of them also for various dopants and over a broad range of doping levels. These reports are surveyed in Table~\ref{Tab:Resonances}.

\begin{table*}[t!]\vspace{-5pt}\footnotesize\centering
\begin{tabular}[c]{@{}l@{~~}c@{~~}l@{~~~}l@{~~}r@{\hspace{1.2pt}}l@{~~}r@{\hspace{1.2pt}}l@{~~~}r@{\hspace{1.2pt}}l@{~~}r@{\hspace{1.2pt}}l@{~~~}r@{}}
\toprule
\multicolumn{2}{l}{$\!$Compound} & sample & $T_{\rm c}$\,(K) & \multicolumn{4}{c}{$\hslash\omega_{\rm res}$ (meV)} & \multicolumn{4}{c}{$\hslash\omega_{\rm res}/k_{\rm B}T_{\rm c}$} & Reference\\
                             & &        &                  & \multicolumn{2}{c}{$\!q_z=0$} & \multicolumn{2}{c}{$\!q_z=\piup$} & \multicolumn{2}{c}{$\!q_z=0$} & \multicolumn{2}{c}{$\!q_z=\piup$} \\
\bottomrule

\multicolumn{10}{l}{$\!$\bf Ba$_{1-x}$K$_x$Fe$_2$As$_2$$^{\strut}$, hole-doped (BKFA)}\\
\midrule
$x=40$\,\% & (OP) & polycryst. & 38 & \multicolumn{2}{c}{---~} & 14.0 & $\pm$\,1.0 & \multicolumn{2}{c}{---~~\,} & 4.3&$\pm$\,0.3 & \citet{ChristiansonGoremychkin08}\\
$x=33$\,\% & ~\hfill»\hfill~ & self-flux  & ~» & 15.0 & $\pm$\,1.0 & 16.0 & $\pm$\,1.0 & 4.6&$\pm$\,0.3 & 4.9&$\pm$\,0.3 & \citet{ZhangWang11}\\
\bottomrule

\multicolumn{10}{l}{$\!$\bf Ba(Fe$_{1-x}$Co$_x$)$_2$As$_2$$^{\strut}$, electron-doped (BFCA)}\\
\midrule
$x=4$\,\% & (UD) & self-flux  & 11 & \multicolumn{2}{c}{---~} & 4.5 & $\pm$\,0.5 & \multicolumn{2}{c}{---~~\,} & 4.7&$\pm$\,0.5 & \citet{ChristiansonLumsden09}\\
$x=4.7$\,\% & ~\hfill»\hfill~ & \quad~\,»  & 17 & 9.0 & $\pm$\,1.0 & 5.0 & $\pm$\,0.5 & 6.1 & $\pm$\,0.7 & 3.4&$\pm$\,0.4 & \citet{PrattKreyssig10}\\
$x=7.5$\,\% & (OP) & \quad~\,»  & 25 & 9.7 & $\pm$\,0.5 & 9.0 & $\pm$\,0.5 & 4.5 & $\pm$\,0.3 & 4.2&$\pm$\,0.3 & \citet{InosovPark10}\\
$x=8$\,\% & (OD) & \quad~\,»  & 22 & 8.6 & $\pm$\,0.5 & 8.6 & $\pm$\,0.5 & 4.5&$\pm$\,0.3 & 4.5&$\pm$\,0.3 & \citet{LumsdenChristianson09}\\
\bottomrule

\multicolumn{10}{l}{$\!$\bf Ba(Fe$_{1-x}$Ni$_x$)$_2$As$_2$$^{\strut}$, electron-doped (BFNA)}\\
\midrule
$x=3.7$\,\% & (UD) & self-flux  & 12.2 & 7.0 & $\pm$\,0.8 & 5.0 & $\pm$\,0.5 & 6.7&$\pm$\,0.8 & 4.8&$\pm$\,0.5 & \citet{WangLuo10}\\
$x=4.5$\,\% & ~\hfill»\hfill~ & \quad~\,»  & 18 & 8.9 & $\pm$\,0.8 & 6.5 & $\pm$\,1.0 & 5.7&$\pm$\,0.5 & 4.2&$\pm$\,0.6 & \citet{ParkInosov10}\\
$x=5$\,\% & (OP) & \quad~\,»  & 20 & 9.1 & $\pm$\,0.4 & 7.2 & $\pm$\,0.5 & 5.3&$\pm$\,0.3 & 4.2&$\pm$\,0.3 & \citet{ChiSchneidewind09}\\
$\phantom{x=}$» & ~\hfill»\hfill~ & \quad~\,»  & ~» & 8.7 & $\pm$\,0.4 & 7.2 & $\pm$\,0.7 & 5.1&$\pm$\,0.3 & 4.2&$\pm$\,0.4 & \citet{LiChen09}\\
$\phantom{x=}$» & ~\hfill»\hfill~ & \quad~\,»  & ~» & 8.0 & $\pm$\,0.5 & \multicolumn{2}{c}{---~~} & 4.6&$\pm$\,0.3 & \multicolumn{2}{c}{---~~\,} & \citet{ZhaoRegnault10}\\
$x=7.5$\,\% & (OD) & \quad~\,»  & 15.5 & 8.0 & $\pm$\,2.0 & 6.0 & $\pm$\,0.5 & 6.0&$\pm$\,1.5 & 4.5&$\pm$\,0.4 & \citet{WangLuo10}\\
\bottomrule

\multicolumn{10}{l}{$\!$\bf Ba(Fe$_{1-x}$Ru$_x$)$_2$As$_2$$^{\strut}$, isovalently substituted (BFRA)}\\
\midrule
$x=25$\,\% & (UD) & self-flux & 14.5 & 8.0 & $\pm$\,1.0 & 5.0 & $\pm$\,1.0 & 6.4 & $\pm$\,0.8 & 4.0 & $\pm$\,0.8 & \citet{ZhaoRotundu13}\\
$x=35$\,\% & (OP) & \quad~\,» & 20 & 8.5 & $\pm$\,1.0 & 7.5 & $\pm$\,1.0 & 4.9 & $\pm$\,0.6 & 4.4 & $\pm$\,0.6 & \citet{ZhaoRotundu13}\\
\bottomrule

\multicolumn{10}{l}{$\!$\bf BaFe$_2$(As$_{1-x}$P$_x$)$_2$$^{\strut}$, isovalently substituted (BFAP)}\\
\midrule
$x=35$\,\% & (OP) & polycryst. & 30 & \multicolumn{2}{c}{---~} & 11.5 & $\pm$\,1.5 & \multicolumn{2}{c}{---~~\,} & 4.5 & $\pm$\,0.6 & \citet{IshikadoNagai11}\\
\bottomrule

\multicolumn{10}{l}{$\!$\bf LaFeAsO$_{1-x}$F$_{x}$$^{\strut}$, electron-doped (La-1111)}\\
\midrule
$x=8$\,\% & (OP) & polycryst. & 29 & \multicolumn{2}{c}{---~} & 13.0 & $\pm$\,1.0 & \multicolumn{2}{c}{---~~\,} & 5.2 & $\pm$\,0.4 & \citet{ShamotoIshikado10}\\
\bottomrule

\multicolumn{10}{l}{$\!$\bf CaFe$_{1-x}$Co$_x$AsF$^{\strut}$, electron-doped (``1111''-type structure)}\\
\midrule
$x=12$\,\% & (OP) & polycryst. & 22 & \multicolumn{2}{c}{---~} & 7.0 & $\pm$\,1.0 & \multicolumn{2}{c}{---~~\,} & 3.7 & $\pm$\,0.5 & \citet{PriceSu13}\\
\bottomrule

\multicolumn{10}{l}{$\!$\bf Li$_{1+\delta}$FeAs$^{\strut}$, undoped (Li-111)}\\
\midrule
\multicolumn{2}{c}{N/A} & polycryst. & 17 & \multicolumn{2}{c}{---~} & 8.0 & $\pm$\,2.0 & \multicolumn{2}{c}{---~~\,} & 5.5 & $\pm$\,1.4 & \citet{TaylorPitcher11}\\
\multicolumn{2}{c}{N/A} & self-flux  & 16.4 & 8.0 & $\pm$\,2.0 & \multicolumn{2}{c}{---~~} & 5.7 & $\pm$\,1.4 & \multicolumn{2}{c}{---~~\,} & \citet{QureshiSteffens12_LiFeAs, QureshiSteffens14}\\
\bottomrule

\multicolumn{10}{l}{$\!$\bf NaFe$_{1-x}$Co$_x$As$^{\strut}$, electron-doped (Na-111)}\\
\midrule
$x=1.5$\,\% & (UD) & self-flux & 15 & 4.5 & $\pm$\,0.5 & 3.25 & $\pm$\,0.5 & 5.3 & $\pm$\,0.4 & 3.5 & $\pm$\,0.4 & \citet{ZhangYu13}\\
$x=4.5$\,\% & (OD) & self-flux & 18 & \multicolumn{2}{c}{---~} & 7.0 & $\pm$\,0.5 & \multicolumn{2}{c}{---~~\,} & 4.5 & $\pm$\,0.3 & \citet{ZhangLi13, ZhangYu13}\\
\bottomrule

\multicolumn{10}{l}{$\!$\bf (CaFe$_{1-x}$Pt$_x$As)$_{10}$Pt$_n$As$_8$$^{\strut}$, platinum-iron-arsenides (``10-4-8'' or ``10-3-8'')}\\
\midrule
$x\approx3$\,\%, $n=4$ & (OD) & single crystal & 30 & 12.5 & $\pm$\,1.0 & \multicolumn{2}{c}{---~~} & 4.8 & $\pm$\,0.4 & \multicolumn{2}{c}{---~~\,} & \citet{SatoKawamata11}\\
$x\approx2$\,\%, $n=4$ & ~\hfill»\hfill~ & single crystal & 33 & 18.0 & $\pm$\,1.5 & \multicolumn{2}{c}{---~~} & 6.3 & $\pm$\,0.5 & \multicolumn{2}{c}{---~~\,} & \citet{IkeuchiSato14}\\
$x\approx6$\,\%, $n=3$ & (OP) & single crystal & 13 & 7.0 & $\pm$\,0.5 & \multicolumn{2}{c}{---~~} & 6.2 & $\pm$\,0.5 & \multicolumn{2}{c}{---~~\,} & \citet{SurmachBrueckner14}\\
\bottomrule

\multicolumn{10}{l}{$\!$\bf FeTe$_{1-x}$Se$_{x}$$^{\strut}$, isovalently substituted (11-family)}\\
\midrule
$x=36$\,\% & (UD) & Bridgman & 12 & 6.5 & $\pm$\,0.5 & \multicolumn{2}{c}{---~} & 6.3 & $\pm$\,0.5 & \multicolumn{2}{c}{---~~\,} & \citet{ChristiansonLumsden13}\\
$x=40$\,\% & (OP) & self-flux & 14 & \multicolumn{2}{c}{---~} & 6.5 & $\pm$\,0.5 & \multicolumn{2}{c}{---~~\,} & 5.3 & $\pm$\,0.4 & \citet{QiuBao09}\\
$\phantom{x=}$» & ~\hfill»\hfill~ & \quad~\,» & ~» & 6.0 & $\pm$\,0.5 & \multicolumn{2}{c}{---~~} & 5.0 & $\pm$\,0.4 & \multicolumn{2}{c}{---~~\,} & \citet{ArgyriouHiess10}\\
$\phantom{x=}$» & ~\hfill»\hfill~ & Bridgman & 13 & 7.0 & $\pm$\,0.5 & \multicolumn{2}{c}{---~} & 6.2 & $\pm$\,0.4 & \multicolumn{2}{c}{---~~\,} & \citet{ChristiansonLumsden13}\\
$x=49$\,\% & ~\hfill»\hfill~ & \quad~\,» & 14 & 7.0 & $\pm$\,0.5 & \multicolumn{2}{c}{---~} & 5.8 & $\pm$\,0.4 & \multicolumn{2}{c}{---~~\,} & \citet{ChristiansonLumsden13}\\
$x=50$\,\% & ~\hfill»\hfill~ & unidirect. solidif.$\!\!$ & ~» & 6.2 & $\pm$\,0.5 & \multicolumn{2}{c}{---~~} & 5.1 & $\pm$\,0.4 & \multicolumn{2}{c}{---~~\,} & \citet{WenXu10}\\
$\phantom{x=}$» & ~\hfill»\hfill~ & Bridgman & ~» & 6.5 & $\pm$\,0.5 & \multicolumn{2}{c}{---~~} & 5.3 & $\pm$\,0.4 & \multicolumn{2}{c}{---~~\,} & \hspace{-1em}\citet{MookLumsden10, MookLumsden10a}, \citet{BabkevichRoessli11}\\
\bottomrule

\multicolumn{10}{l}{$\!$\bf \textit{A}$_x$Fe$_{2-y}$Se$_2$$^{\strut}$, alkali-metal iron selenides (245-family)}\\
\midrule
$A$~=~K & (OP) & float.~zone & 32 & 14 & $\pm$\,1.0 & \multicolumn{2}{c}{---~~} & 5.1 & $\pm$\,0.4 & \multicolumn{2}{c}{---~~\,} & \citet{FriemelLiu12}\\
$A$~=~Rb & ~\hfill»\hfill~ & Bridgman & 32 & 14 & $\pm$\,1.0 & \multicolumn{2}{c}{---~~} & 5.1 & $\pm$\,0.4 & \multicolumn{2}{c}{---~~\,} & \citet{ParkFriemel11}\\
$\phantom{x=}$» & ~\hfill»\hfill~ & \quad~\,» & ~» & 14 & $\pm$\,1.0 & \multicolumn{2}{c}{---~~} & 5.1 & $\pm$\,0.4 & \multicolumn{2}{c}{---~~\,} & \citet{FriemelPark12}\\
$A$~=~Cs & ~\hfill»\hfill~ & \quad~\,» & 27 & 11 & $\pm$\,1.0 & \multicolumn{2}{c}{---~~} & 4.7 & $\pm$\,0.4 & \multicolumn{2}{c}{---~~\,} & \citet{TaylorEwings12}\\
\multicolumn{2}{l}{\hspace{-6pt}$A$~=~Li$_x$(ND$_2$)$_y$(ND$_3$)$_{1-y}$} & polycryst.$\!\!$ & 43 & 19 & $\pm$\,3 & \multicolumn{2}{c}{---~~} & 5.0 & $\pm$\,1.0 & \multicolumn{2}{c}{---~~\,} & \citet{TaylorSedlmaier13}\\
\bottomrule
\end{tabular}
\caption{Summary of the spin resonance energies, $\omega_{\rm res}$, and the corresponding $\omega_{\rm res}/k_{\rm B}T_{\rm c}$ ratios in iron-based superconductors.\label{Tab:Resonances}}
\vspace*{-3em}
\end{table*}

\begin{figure*}[b!]\centering
\includegraphics[width=0.95\textwidth]{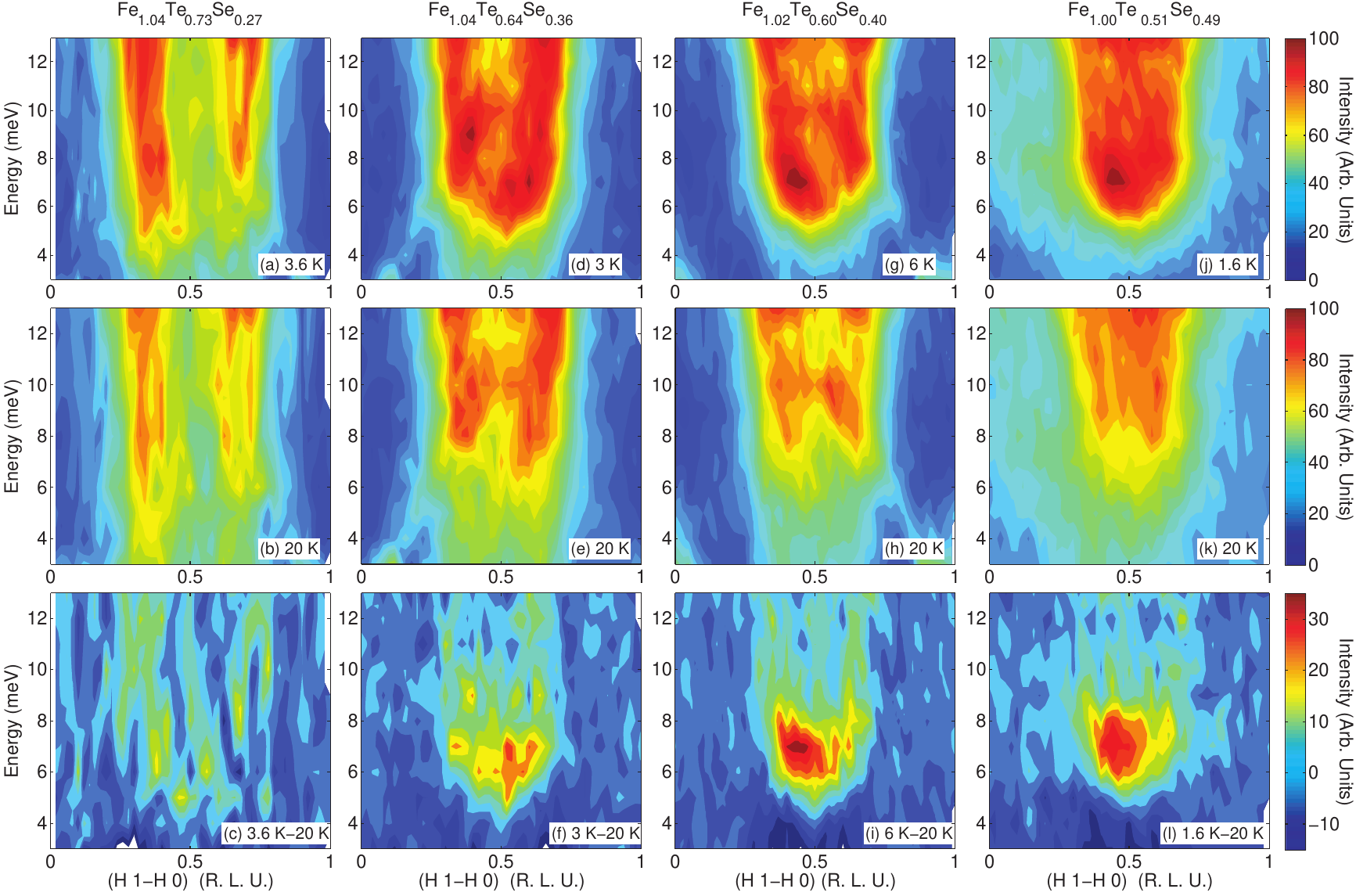}
\caption{Evolution of the low-temperature (top row) and normal-state (middle row) spectra in Fe$_{1+y}$Te$_{1-x}$Se$_x$, along with the resonant intensity, represented by their difference (bottom row), as a function of the Se doping level. The momentum-space notation corresponds to two Fe atoms per unit cell, in which the $(H\,1\!-\!H\,0)$ direction of the presented cuts is equivalent to the $(\half\,K\,0)$ direction in the unfolded-zone notation. Reproduced with permission from Ref.\,\citenum{ChristiansonLumsden13}, copyright by the American Physical Society.}
\label{Fig:FeTeSe-resonance}
\end{figure*}

In order not to reiterate the results summarized already in previous reviews \cite{Johnston10, LumsdenChristianson10review, Stewart11, InosovPark11}, I would like to focus here only on the most recent publications and the unusual behavior of the resonant mode in several novel materials. Among the most studied ``122''-type compounds, the conventional resonant mode has been investigated recently in the isovalently substituted Ba-122 samples, such as polycrystalline BaFe$_2$(As$_{1-x}$P$_x$)$_2$ \cite{IshikadoNagai11} and single-crystalline Ba(Fe$_{1-x}$Ru$_x$)$_2$As$_2$ \cite{ZhaoRotundu13}. A clear redistribution of spectral weight below $T_{\rm c}$ could be also seen in polycrystalline ``1111''-type pnictides, such as LaFeAsO$_{0.92}$F$_{0.08}$ \cite{ShamotoIshikado10} and CaFe$_{0.88}$Co$_{0.12}$AsF \cite{PriceSu13}. Of particular interest is the anomalous behavior of the magnetic resonance in the ``10-4-8'' \cite{SatoKawamata11, IkeuchiSato14} and ``10-3-8'' \cite{SurmachBrueckner14} iron-platinum-arsenide compounds. While resonance-like peaks have been observed in these systems at the usual $(\piup,0)$ wave vector, their unusually large $\hslash\omega_{\rm res}/k_{\rm B}T_{\rm c}$ ratios (see Table\,\ref{Tab:Resonances}) and their persistence above $T_{\rm c}$ are not typical for other iron-based superconductors. This unconventional behavior was recently explained by the appearance of a pseudogap and possible preformed-pair formation evidenced by a drop of the spin-lattice relaxation rate in the NMR signal at low temperatures \cite{SurmachBrueckner14}.

In the ``11''-type iron-selenide and iron-telluride superconductors, the momentum-space structure of spin fluctuations qualitatively differs from that of most iron pnictides. The E-type bicollinear AFM ordering pattern in the Fe$_{1+x}$Te parent compound results in incommensurate or stripy excitations with an ``hourglass''-like dispersion stemming from the $(\quarter \quarter)$ propagation vector \cite{ZaliznyakXu11, StockRodriguez14}. With increasing interstitial iron concentration, these excitations soften, become gapless, and shift to an incommensurate position \cite{StockRodriguez11, StockRodriguez14}. Upon moderate Se doping, the higher-energy part of the spectrum is further displaced towards the $(\half 0)$ wave vector \cite{ChiRodriguezRivera11}, and depending on the excess iron content, a constriction toward a commensurate ``neck'' of an hourglass-shaped dispersion develops in superconducting samples above $T_{\rm c}$, subsequently resulting in a commensurate resonance peak centered at the usual $(\half 0)$ wave vector \cite{QiuBao09, BabkevichBendele10, BabkevichRoessli11, TsyrulinViennois12}. This spectacular evolution from incommensurate spin excitations towards the commensurate resonance as a function of Se doping, most clearly visualized in Ref.~\citenum{ChristiansonLumsden13}, is presented below in Fig.\,\ref{Fig:FeTeSe-resonance}.

\begin{figure*}[t!]
\hspace{-0.005\textwidth}\includegraphics[width=1.01\textwidth]{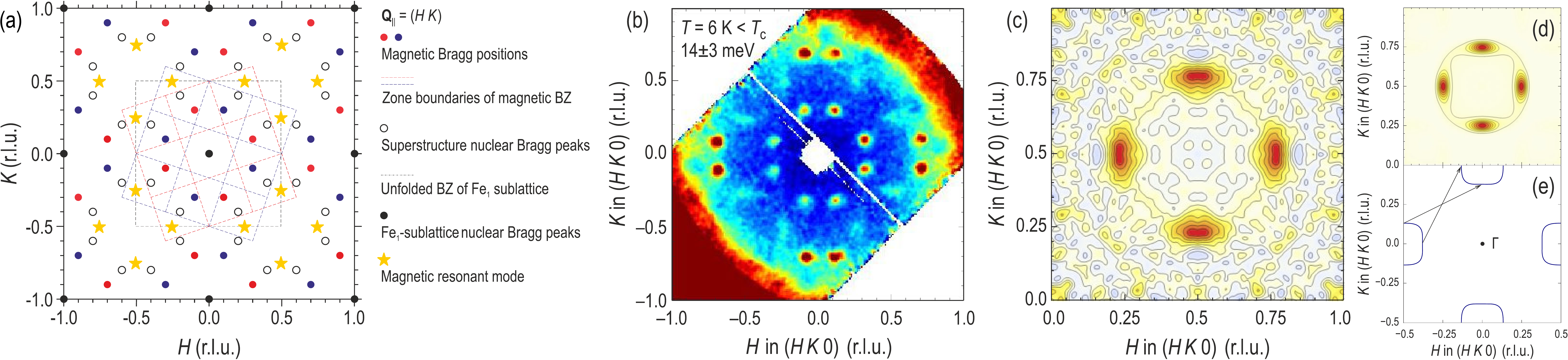}
\caption{(a)~The in-plane projection of the magnetic and nuclear Bragg peak positions arising from the $\sqrt{5} \times\! \sqrt{5}$ iron-vacancy superstructure and the positions of resonant-mode excitations in phase-separated $A_x$Fe$_{2-y}$Se$_2$ superconductors, adapted from Ref.\,\citenum{ParkFriemel11}. (b)~A constant-energy cut through the low-temperature TOF data on K$_x$Fe$_{2-y}$Se$_2$, integrated around the resonance energy of 14\,meV. The sharp spots represent spin-wave excitations from the antiferromagnetic insulating phase, whereas diffuse elongated features originate from the magnetic resonance intensity in the superconducting phase \cite{FriemelLiu12}. (c)~Resonant intensity distribution in Rb$_x$Fe$_{2-y}$Se$_2$, obtained by interpolating the difference of the low-temperature and normal-state INS intensities at 15\,meV, as described in Ref.\,\citenum{FriemelPark12}. (d)~The corresponding difference of the calculated imaginary parts of the dynamic spin susceptibility for the superconducting and normal states, taken at the resonance energy. The calculation was done within the RPA from the tight-binding band model of \textit{A}$_x$Fe$_2$Se$_2$ \cite{MaierGraser11}, which was rigidly shifted to match the experimental peak positions. (e) The resulting Fermi surface in the $(H\,K\,0)$ plane with black arrows indicating the in-plane nesting vectors responsible for the peaks of intensity observed in panel (d). Copyright by the American Physical Society.}
\label{Fig:RFS-resonance}\vspace{-1pt}
\end{figure*}

In the related ``245'' family of alkali-metal iron selenides, $A_x$Fe$_{2-y}$Se$_2$ ($A$\,=\,K, Rb, Cs, Tl), the situation is further complicated by the structural phase separation between vacancy-ordered insulating antiferromagnetic and vacancy-free superconducting phases, discussed already in section \ref{SubSec:IronChalcogenides}. As a result, the magnetic INS response of such samples consists of two distinct contributions: (i) eight three-dimensional conical branches of acoustic spin-wave excitations from two antiferromagnetic twin domains with different orientation of the iron-vacancy superstructure \cite{XiaoNandi13, ChiYe13, Bao15} and (ii) an itinerant two-dimensional magnetic response of the vacancy-free metallic phase, exhibiting the magnetic resonant mode below $T_{\rm c}$ \cite{ParkFriemel11, FriemelPark12, FriemelLiu12, TaylorEwings12}. The two contributions reside in different places in the reciprocal space, as schematically shown in Fig.\,\ref{Fig:RFS-resonance}\,(a), which allows us to distinguish them clearly in INS experiments performed on single crystals, see for instance Fig.\,\ref{Fig:RFS-resonance}\,(b). Unlike in most other families of iron-based superconductors discussed above, in $A_x$Fe$_{2-y}$Se$_2$ the normal-state paramagnon intensity and, consequently, the magnetic resonant mode are found at the $(\half,\pm\quarter)$ and equivalent wave vectors \cite{ParkFriemel11, FriemelPark12, TaylorEwings12}. This is illustrated by the experimentally measured difference of the superconducting- and normal-state INS intensities in Fig.\,\ref{Fig:RFS-resonance}\,(c). From the theory point of view, this marked difference to other ferropnictides can be explained by the absence of the $\Gamma$-centered hole pocket in the highly electron-doped band structure of alkali-metal iron selenides, which leads to qualitatively different nesting vectors connecting the sides of two electron pockets, as shown in Fig.\,\ref{Fig:RFS-resonance}\,(d,e). However, within this scenario the wave vector of the excitations should be strongly doping dependent, and it still remains unclear why it is pinned to the commensurate $(\half,\pm\quarter)$ position in all samples studied this far \cite{FriemelLiu12}. It is likely that future experiments on differently intercalated iron selenides \cite{TaylorSedlmaier13} may shed light on this problem.

Incommensurate normal-state excitations close to the $(\piup,0)$ point have been observed in LiFeAs \cite{TaylorPitcher11, WangWang12, QureshiSteffens12_LiFeAs, QureshiSteffens14}, experiencing only a relatively weak spectral-weight redistribution below $T_{\rm c}$ that was employed among other arguments \cite{KnolleZabolotnyy12} to support alternative pairing symmetries in LiFeAs, such as $s_{++}$ \cite{SaitoOnari14}, $p$-wave \cite{BrydonDaghofer11, BaekHarnagea13}, orbital-antiphase $s_\pm$ \cite{YinHaule14}, $s\!+\!{\rm i}d$ \cite{HankeSykora12}, or even more exotic chiral order parameters \cite{LiUrbano13}. While heated discussions about the order-parameter symmetry in LiFeAs continue, it is useful to understand the possible implications of a resonant mode observed by neutron scattering for the gap structure. It is often assumed that the formation of a pronounced resonant mode necessarily requires a sign-changing order parameter \cite{KorshunovEremin08, MaierScalapino08, MaierGraser09}, which is further corroborated by the observation of strong neutron resonances in $d$-wave copper oxides and $s_{\pm}$-wave iron pnictides. However, some iron-based superconductors like LiFeAs exhibit only a weak redistribution of spectral weight below $T_{\rm c}$ that does not lead to any visible peak in the raw INS spectrum and requires a careful comparison with the normal-state intensity to be observed. Within an itinerant-electron model, such weak effects are expected even for the conventional $s$-wave pairing due to the depletion of the low-energy spectral weight within the gap region and consequent formation of a ``pile-up peak'' above $2\Delta$. It has been argued using model calculations that the INS peak obtained in the $s_{++}$ state can resemble the experimentally observed magnetic resonant mode \cite{OnariKontani10, OnariKontani11}. There are several complications on the way to resolving this argument from both experimental and theoretical sides:\vspace{-2pt}

\begin{figure*}[b!]\vspace{-10pt}
\begin{minipage}[b]{0.49\textwidth}
\includegraphics[width=\textwidth]{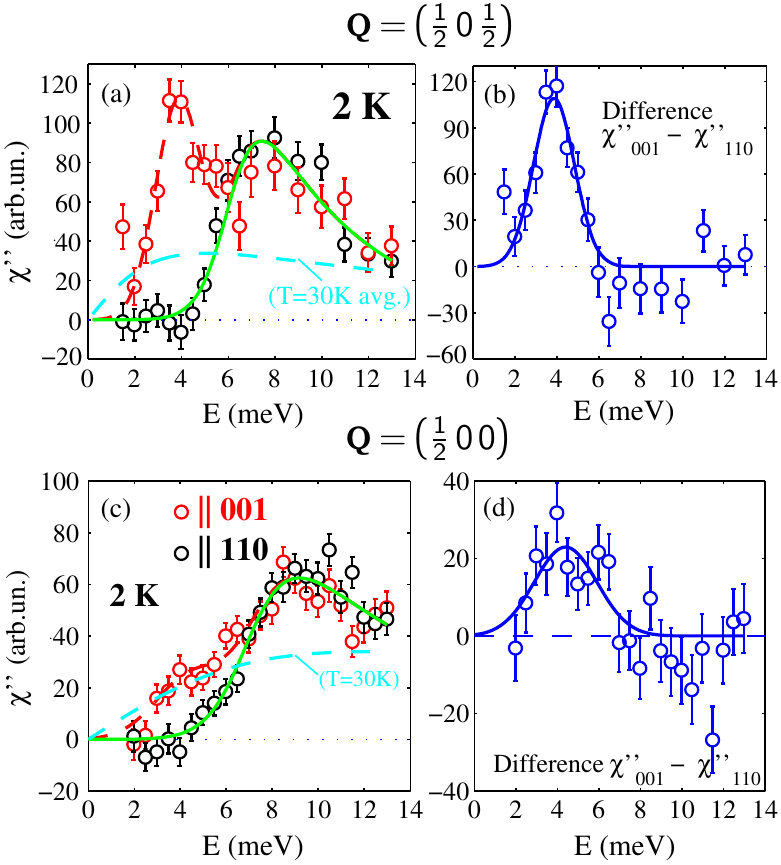}
\end{minipage}\hfill
\begin{minipage}[b]{0.49\textwidth}
\caption{\footnotesize The two-component structure and spin anisotropy of the resonant mode in Ba(Fe$_{0.94}$Co$_{0.06}$)$_2$As$_2$. (a,c)~Energy dependence of the imaginary
part of the generalized magnetic susceptibility at $\mathbf{Q}=(\half\,0\,0)$ and $(\half\,0\,\half)$, respectively, for the out-of-plane and in-plane spin polarization components. (b,d)~The corresponding differences of the out-of-plane and in-plane polarization channels, singling out the lower-energy anisotropic resonant contribution. Reproduced from Ref.\,\citenum{SteffensLee13}, copyright by the American Physical Society.\vspace{12pt}}
\label{Fig:ResAnisotropy}
\end{minipage}\vspace{-10pt}
\end{figure*}

\begin{enumerate}
\item \emph{Absolute spectral weight.} It is difficult to define and compare the strength of the resonant mode (especially among different compounds), as it requires both INS measurements and theoretical calculations to be performed in absolute intensity units. Moreover, for any gap symmetry, the resonance peak is formed from the normal-state spectral weight collected within an energy window of the order of $2\Delta$, and is therefore much more dependent on the low-energy spectral weight in the normal state than on the superconducting order parameter. Because the true resonance mode in the case of a sign-changing superconducting gap is expected to appear below $2\Delta$, its integrated spectral weight can be even lower in comparison to a pile-up peak in the $s$-wave case, which collects the spectral weight from the whole $2\Delta$ energy range.
\item \emph{Spectral-weight conservation.} The sum rule for the scattering function requires that the total momentum- and energy-integrated magnetic spectral weight remains constant as a function of temperature. Hence, a pile-up peak must unavoidably form above $2\Delta$ as a consequence of spin-gap opening. However, RPA-based models are not expected to be restricted by the spectral-weight conservation, because they are reliable only in the low-energy limit, whereas the total spectral weight is integrated over the entire energy range. As a result, direct comparison of the RPA calculations for the normal and superconducting states can be misleading, resulting in a nonphysical situation when the calculated normal-state response stays always below the superconducting response calculated for the $s$-wave symmetry at all energies \cite{KnolleZabolotnyy12}. Rescaling one of the curves to satisfy the sum rule below a certain threshold could possibly cure this obvious problem at the expense of introducing an additional tuning parameter in the model.
\item \emph{Energy of the resonance.} It is in principle possible to distinguish a true collective resonant excitation from a pile-up peak by comparing its energy with the value of the superconducting gap, $2\Delta$. However, in iron-based superconductors this analysis is complicated by the presence of multiple gaps \cite{InosovPark11} and a relatively broad resonant mode consisting of several contributions from different spin-polarization channels \cite{SteffensLee13, ZhangYu13}. Furthermore, the accuracy of the available gap measurements is often comparable both with the width of the peak and with the separation between the resonant mode and $2\Delta$ \cite{InosovPark10, InosovPark11}, leaving a degree of uncertainty in the relative placement of the two energy scales.
\item \emph{Width of the resonance.} It is also expected that a collective resonant mode below $2\Delta$ should be much sharper due to the absence of scattering from the particle-hole continuum. In contrast, a pile-up peak above the gap edge would extend over an energy range of the order of $2\Delta$. However, numerous alternative sources of broadening (contribution from different polarization components, $k_z$ dispersion, impurity scattering, etc.) may smear this difference and lead to rather broad resonant signals \cite{InosovPark10}.\vspace{-2pt}
\end{enumerate}

As a result, there appears to be no straightforward criterion that would allow us to distinguish between different pairing symmetries based exclusively on the presence of the neutron spin resonance. Convincing arguments are only possible using accurate quantitative comparisons of the calculated and measured resonant spectral weights, peak shapes, and $\omega_{\rm res}/2\Delta$ ratios, in combination with complementary experimental methods.

\subsection{Spin anisotropy of the resonant mode}

It has been recently ascertained that in some compounds, in particular in electron-underdoped NaFe$_{0.985}$Co$_{0.015}$As \cite{ZhangYu13} and nearly optimally doped Ba(Fe$_{0.94}$Co$_{0.06}$)$_2$As$_2$ \cite{SteffensLee13}, the magnetic resonant mode is split into two peaks separated by $\sim$\,3\,meV in energy, as shown in Fig.\,\ref{Fig:ResAnisotropy}. The sharper lower-energy mode showed a significant dispersion along the $Q_z$ direction, while the broader higher-energy mode was dispersionless. The splitting vanished at higher doping levels, so that in the overdoped NaFe$_{0.935}$Co$_{0.045}$As, only a single resonance could be observed \cite{ZhangYu13}. Neutron polarization analysis revealed that the lower-energy resonant mode has a well pronounced spin anisotropy with predominant out-of-plane polarization, while the higher-energy mode was nearly isotropic \cite{SteffensLee13, ZhangSong14}. This
\begin{wrapfigure}[24]{r}{0.4\textwidth}
\noindent\includegraphics[width=0.4\textwidth]{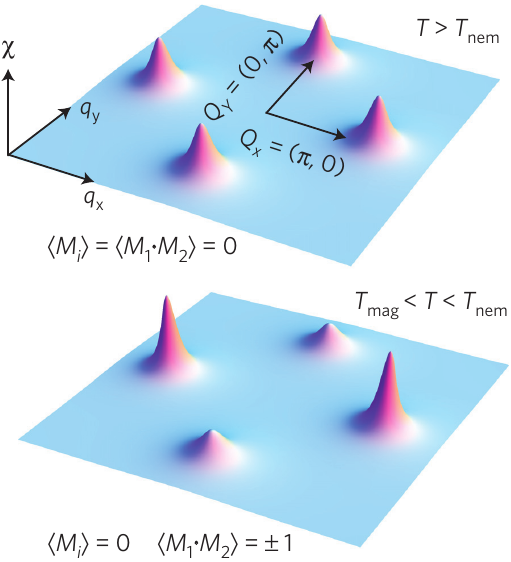}\vspace{-3.5pt}
\caption{Schematically illustrated momentum dependence of the dynamic spin susceptibility in the tetragonal paramagnetic phase (top) and in the orthorhombic spin-nematic phase (bottom), which results in unequal intensities of the $(\piup,0)$ and $(0,\piup)$ paramagnon branches. Reproduced with permission from Ref.\,\citenum{FernandesChubukov14}.}
\label{Fig:NematicTheory}
\end{wrapfigure}
behavior mimics the anisotropy of spin-wave excitations in the corresponding parent compounds, where the $c$-axis-polarized magnons are characterized by the smallest spin gap \cite{QureshiSteffens12, SongRegnault13}.\enlargethispage{5pt}

A qualitatively similar anisotropy of the neutron spin resonance has also been observed in many other iron-based superconductors, even when the two contributions could not be resolved as separate peaks in a non-polarized INS spectrum. For instance, a comparison of the nearly optimally doped ($x=0.05$) and overdoped ($x=0.075$) Ba(Fe$_{1-x}$Ni$_x$)$_2$As$_2$ revealed that the anisotropy persists only in proximity to the antiferromagnetic parent compound and vanishes in overdoped samples \cite{LipscombeHarriger10, LiuLester12}. In the weakly underdoped regime, where superconductivity coexists with the static AFM order ($x=0.048$, $T_{\rm N}\approx T_{\rm s}=33$\,K, $T_{\rm c}=19.8$\,K), pronounced spin anisotropy could be observed even in the paramagnetic state far above $T_{\rm N}$, and was even further enhanced in the superconducting state \cite{LuoWang13}. Similar effects have been observed in the hole-doped Ba$_{1-x}$K$_x$Fe$_2$As$_2$ \cite{ZhangLiu13} with the substantial difference that the anisotropy persists into the overdoped regime far beyond the suppression of magnetic order, even when the low-energy magnetic excitations become incommensurate \cite{QureshiLee14}. In the same overdoped sample ($x=0.5$), the resonance energy exhibits a non-negligible dispersion along the $Q_z$ direction, shifting from 14.7\,meV at the magnetic zone center to 15.7\,meV at the zone boundary.

In the ``11''-type FeSe$_{0.5}$Te$_{0.5}$, the in-plane and out-of-plane components of the magnetic resonant mode despite their somewhat different intensities are peaked at the same energy \cite{BabkevichRoessli11}, in contrast to the ``122'' compounds. The in-plane polarization channel provides the dominant contribution to the resonant mode intensity. Moreover, spin fluctuations remain nearly isotropic both above and below the resonance peak \cite{BabkevichRoessli11}. A similarly weak spin-space anisotropy with a slightly stronger out-of-plane component was also observed recently for the incommensurate magnetic fluctuations in LiFeAs around the resonance energy between 8 and 10~meV \cite{QureshiSteffens14}.

The anisotropy of the resonant mode challenges its simplified understanding as an isotropic triplet excited state of the singlet Cooper pairs \cite{LipscombeHarriger10}. On the one hand, it emphasizes the orbital dependence of the superconducting pairing in iron pnictides \cite{ZhangYu13} and a close relationship of the collective spin fluctuations within the superconducting phase with the spin waves of the antiferromagnetic parent compounds \cite{SteffensLee13}. It has been therefore suggested that the spin excitation anisotropy could be probing spontaneously broken electronic symmetries such as orbital ordering in the tetragonal phase of iron pnictides \cite{LuoWang13}. On the other hand, a recent theoretical work \cite{LvMoreo14} suggested that the coexistence of either static or fluctuating magnetic order with superconductivity could also have a strong effect on the resonant mode and lead to distinct resonance energies at $(0,\piup)$ and $(\piup,0)$ wave vectors, which would result in a two-peak structure in twinned samples, similar to the experimental observations \cite{ZhangYu13, SteffensLee13}. Which of the two mechanisms (orbital or magnetic) is responsible for the complex structure of the resonant mode in ferropnictides can possibly be distinguished by future INS experiments on detwinned superconducting samples under a uniaxial pressure.

\begin{wrapfigure}[30]{r}{0.51\textwidth}\vspace{-67pt}
\noindent\includegraphics[width=0.51\textwidth]{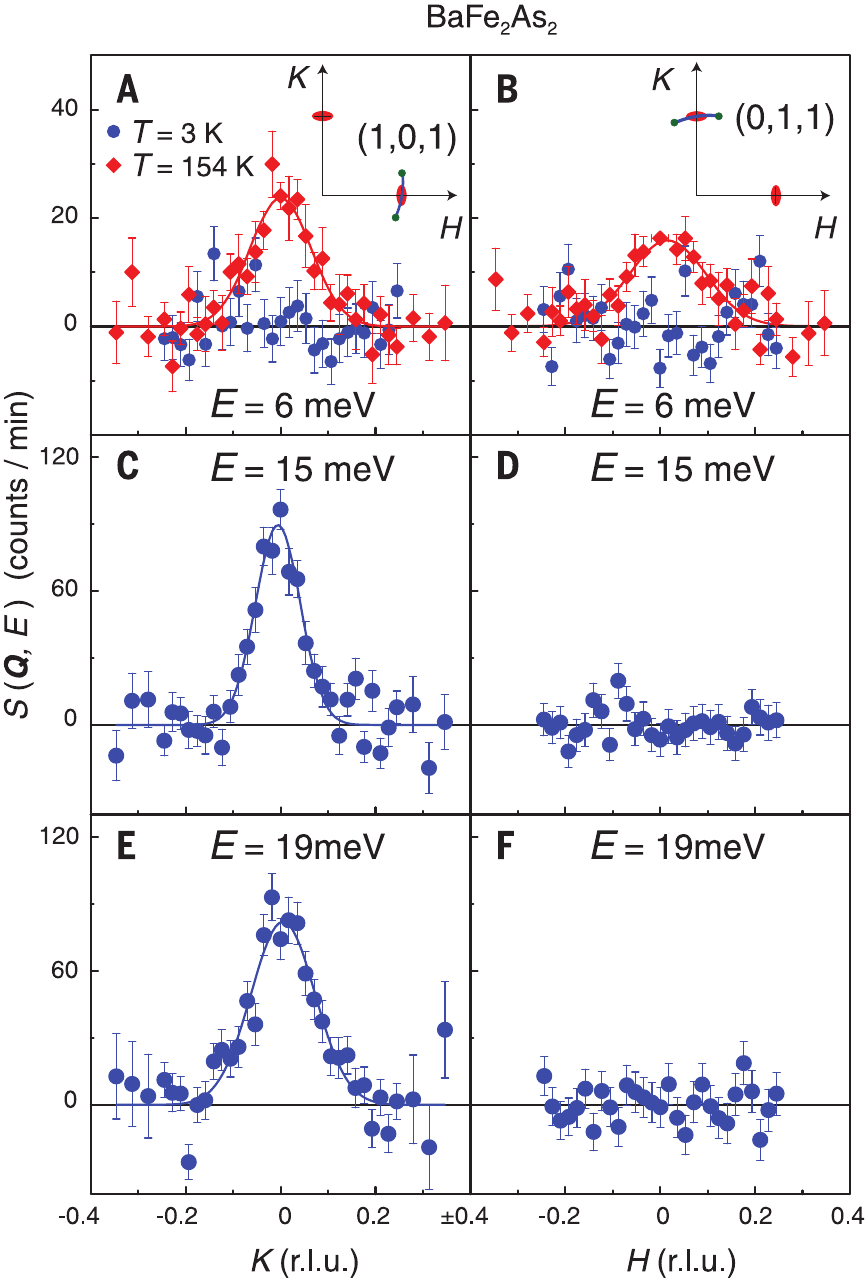}\vspace{-2pt}
\caption{Low-energy spin-wave intensity in a single crystal of BaFe$_2$As$_2$ detwinned by a uniaxial pressure. The complete suppression of inelastic scattering below $T_{\rm N}$ is seen at the $(0,\piup)$ wave vector (right column), while the $(\piup,0)$ scattering is enhanced in comparison to the paramagnetic state (left column). Reproduced with permission from Ref.\,\citenum{LuPark14}.}\label{Fig:BFA-detwinned}
\end{wrapfigure}

\section{Magnetic dynamics in the spin-nematic state}

\subsection{Theoretical and experimental progress}

Finally, let us consider the spectrum of spin fluctuations in the spin-nematic state, i.e. within the orthorhombic paramagnetic region spanning $T_{\rm s}$ and $T_{\rm N}$ in the generic phase diagram. Due to the necessity to perform experiments on detwinned single crystals and the numerous technical complications associated with this procedure, described already in section~\ref{SubSec:UnderstandingNematic}, this phase still remains nearly unexplored by neutron spectroscopy. According to theoretical expectations \cite{FernandesChubukov14}, the spin-nematic state should break the equivalency of the two excitation branches at $(\piup,0)$ and $(0,\piup)$, leading to a disproportionation of the inelastic scattering intensities between the two wave vectors, as schematically illustrated in Fig.\,\ref{Fig:NematicTheory}. Upon cooling, the transfer of spectral weight from $(0,\piup)$ to $(\piup,0)$ should start already at $T_{\rm s}$ and be complete upon reaching $T_{\rm N}$, where the $(0,\piup)$ fluctuations were experimentally shown to be fully gapped \cite{LuPark14}. This understanding is generally consistent with the progressive development of the twofold anisotropy seen in NMR measurements on LaFeAsO \cite{FuTorchetti12}. Still, this plausible but simplified description does not provide specific predictions as to the exact spectrum of spin fluctuations as a function of energy and temperature. While the $(0,\piup)$ branch is expected to lose all of its intensity up to $\sim$\,200\,meV below $T_{\rm N}$, it is doubtful that this whole energy range could be equally affected immediately below $T_{\rm s}$. One likely scenario would imply that the changes begin only at low energies and then propagate towards the top of the magnon band only as the temperature is decreased. Further, the expected temperature dependence of the anisotropy gap at low energies still remains unaddressed by the theory \cite{ParkFriemel12}, as it would require more involved multiband calculations with the full account of the spin-orbit coupling effects.

Experimentally, a pioneering INS study on a detwinned single crystal of BaFe$_2$As$_2$ measured under constant uniaxial pressure has been performed recently by X.~Lu \textit{et al.} \cite{LuPark14}. Using a specially designed detwinning device for large crystals, they could measure for the first time the spin-wave spectrum of the Ba-122 parent compound and demonstrated the complete absence of scattering intensity at the $(0,\piup)$ wave vector below $T_{\rm N}$, as illustrated in Fig.\,\ref{Fig:BFA-detwinned}. However, because the structural and magnetic phase transitions in BaFe$_2$As$_2$ coincide, the direct visualization of the spectrum within the spin-nematic phase, like the one depicted in Fig.\,\ref{Fig:NematicTheory}, can not be achieved unless a different compound with a large difference of $T_{\rm s}$ and $T_{\rm N}$ is used instead.

So far, experiments of this kind were performed only on twinned single crystals of Ba(Fe$_{0.953}$Co$_{0.047}$)$_2$As$_2$ and LaFeAsO by Q.~Zhang \textit{et al.} without the application of uniaxial pressure \cite{ZhangFernandes15}. They revealed a significant feedback effect of the nematic order on the low-energy magnetic spectrum, manifested in a sharp nearly threefold decrease of the zone-center line width in momentum space with an onset at $T_{\rm s}$ that reached its minimum around $T_{\rm N}$ and then stayed nearly constant at even lower temperatures. According to the authors, the strong enhancement of the spin-spin correlation length implied by this observation is caused by the coupling of the two fluctuating N\'{e}el sublattices via the orthorhombic lattice distortion below $T_{\rm s}$. They propose a theoretical model for the correlation length that takes into account the presence of two symmetry-related magnetic instabilities, giving rise to the preemptive spin-nematic transition at $T_{\rm s} > T_{\rm N}$, and ultimately shows good qualitative agreement with the experimental data \cite{ZhangFernandes15}. The neutron-scattering results are consistent with the earlier $^{75}$As NMR evidence for the enhancement of spin fluctuations within the spin-nematic phase, manifested in the rapid increase of the spin-lattice relaxation rate just below $T_{\rm s}$ \cite{MaChen11}.

\subsection{Outlook}

In view of the ever increasing interest to the spin-nematic state of iron pnictides and the remaining open questions regarding its origin and relationship to unconventional superconductivity, reports from various new exciting experiments are to be expected in the next years. Among them, neutron-scattering studies under small uniaxial pressure applied \textit{in situ} would definitely play a central role. Such experiments would require certain technical developments for the creation of a sample environment capable of simultaneously satisfying all the conditions listed in section~\ref{SubSec:UnderstandingNematic}. It should allow for the reproducible, uniform, well controlled application and release of the uniaxial stress at low temperatures, provide access to both $(\piup,0)$ and $(0,\piup)$ wave vectors without changing the sample configuration, offer the applicability to single crystals of moderate sizes or coaligned crystal assemblies, and still meet the usual requirements with respect to the dimensional constraints and low neutron-scattering background for compatibility with existing cryostats. Any technical progress in this direction is expected to have a profound impact extending beyond the field of iron-based superconductors, as similar experiments can be then attempted on other materials in the proximity to structural instabilities.

As already mentioned, it would be particularly interesting to establish experimentally the spectrum of spin fluctuations within the spin-nematic phase on detwinned single crystals of different iron-pnictide parent compounds after the \textit{in situ} stress release and to study its temperature-, doping- and strain-dependencies. It is also worthwhile to perform polarization analysis of the excitations within the single-domain magnetic state. Furthermore, similar experiments can be carried out on superconducting samples to establish the influence of uniaxial stress on the magnetic resonant mode below $T_{\rm c}$ and on its spin-space anisotropy.\vspace{-2pt}

\section*{Acknowledgements}

The author acknowledges financial support from DFG within the priority program SPP~1458, under Grant No. IN~209/1-2, and the research training group GRK~1621 at the TU Dresden.\vspace{-2pt}


%

\end{document}